\documentclass[twocolumn,
              superscriptaddress,
              prc,
              showkeys,
              showpacs,
              nofootinbib,
              notitlepage,
              floatfix,
              ]{revtex4-2}

\usepackage[perpage]{footmisc}

\usepackage{graphicx}
\usepackage[colorlinks = true,
            linkcolor = purple,
            urlcolor  = blue,
            citecolor = purple,
            anchorcolor = blue]{hyperref}
\usepackage{xcolor}
\usepackage{amssymb}
\usepackage[nointegrals]{wasysym}
\usepackage{amsthm}
\usepackage{amsmath}
\usepackage{textcomp}
\usepackage{mathtools}
\usepackage{multirow}
\usepackage{upgreek}
\usepackage{isotope}
\usepackage{float}
\usepackage{soul, ulem}

\usepackage{lineno}

\usepackage[separate-uncertainty,retain-explicit-plus,per-mode = symbol]{siunitx}
\usepackage{url}
\usepackage{array}
\usepackage{nicefrac}

\renewcommand{\arraystretch}{1.5}

\usepackage{booktabs}

\newcommand{\doublebeta}{\ensuremath{0\upnu\upbeta\upbeta}}
\newcommand{\twonubeta}{\ensuremath{2\upnu\upbeta\upbeta}}
\newcommand{\dec}{\ensuremath{2\upnu\text{ECEC}}}

\newcommand{\nt}{XENONnT}
\newcommand{\mbb}{\ensuremath{\langle m_{\upbeta\upbeta} \rangle}}

\newcommand{\bologna}{\affiliation{Department of Physics and Astronomy, University of Bologna and INFN-Bologna, 40126 Bologna, Italy}}
\newcommand{\chicago}{\affiliation{Department of Physics \& Kavli Institute for Cosmological Physics, University of Chicago, Chicago, IL 60637, USA}}
\newcommand{\coimbra}{\affiliation{LIBPhys, Department of Physics, University of Coimbra, 3004-516 Coimbra, Portugal}}
\newcommand{\columbia}{\affiliation{Physics Department, Columbia University, New York, NY 10027, USA}}
\newcommand{\lngs}{\affiliation{INFN-Laboratori Nazionali del Gran Sasso and Gran Sasso Science Institute, 67100 L'Aquila, Italy}}
\newcommand{\mainz}{\affiliation{Institut f\"ur Physik \& Exzellenzcluster PRISMA$^{+}$, Johannes Gutenberg-Universit\"at Mainz, 55099 Mainz, Germany}}
\newcommand{\heidelberg}{\affiliation{Max-Planck-Institut f\"ur Kernphysik, 69117 Heidelberg, Germany}}
\newcommand{\munster}{\affiliation{Institut f\"ur Kernphysik, Westf\"alische Wilhelms-Universit\"at M\"unster, 48149 M\"unster, Germany}}
\newcommand{\nikhef}{\affiliation{Nikhef and the University of Amsterdam, Science Park, 1098XG Amsterdam, Netherlands}}
\newcommand{\nyuad}{\affiliation{New York University Abu Dhabi - Center for Astro, Particle and Planetary Physics, Abu Dhabi, United Arab Emirates}}
\newcommand{\purdue}{\affiliation{Department of Physics and Astronomy, Purdue University, West Lafayette, IN 47907, USA}}
\newcommand{\rice}{\affiliation{Department of Physics and Astronomy, Rice University, Houston, TX 77005, USA}}
\newcommand{\stockholm}{\affiliation{Oskar Klein Centre, Department of Physics, Stockholm University, AlbaNova, Stockholm SE-10691, Sweden}}
\newcommand{\subatech}{\affiliation{SUBATECH, IMT Atlantique, CNRS/IN2P3, Universit\'e de Nantes, Nantes 44307, France}}
\newcommand{\torino}{\affiliation{INAF-Astrophysical Observatory of Torino, Department of Physics, University  of  Torino and  INFN-Torino,  10125  Torino,  Italy}}
\newcommand{\ucsd}{\affiliation{Department of Physics, University of California San Diego, La Jolla, CA 92093, USA}}
\newcommand{\wis}{\affiliation{Department of Particle Physics and Astrophysics, Weizmann Institute of Science, Rehovot 7610001, Israel}}
\newcommand{\zurich}{\affiliation{Physik-Institut, University of Z\"urich, 8057  Z\"urich, Switzerland}}
\newcommand{\paris}{\affiliation{LPNHE, Sorbonne Universit\'{e}, CNRS/IN2P3, 75005 Paris, France}}
\newcommand{\freiburg}{\affiliation{Physikalisches Institut, Universit\"at Freiburg, 79104 Freiburg, Germany}}
\newcommand{\napels}{\affiliation{Department of Physics ``Ettore Pancini'', University of Napoli and INFN-Napoli, 80126 Napoli, Italy}}
\newcommand{\nagoya}{\affiliation{Kobayashi-Maskawa Institute for the Origin of Particles and the Universe, and Institute for Space-Earth Environmental Research, Nagoya University, Furo-cho, Chikusa-ku, Nagoya, Aichi 464-8602, Japan}}
\newcommand{\laquila}{\affiliation{Department of Physics and Chemistry, University of L'Aquila, 67100 L'Aquila, Italy}}
\newcommand{\tokyo}{\affiliation{Kamioka Observatory, Institute for Cosmic Ray Research, and Kavli Institute for the Physics and Mathematics of the Universe (WPI), University of Tokyo, Higashi-Mozumi, Kamioka, Hida, Gifu 506-1205, Japan}}
\newcommand{\kobe}{\affiliation{Department of Physics, Kobe University, Kobe, Hyogo 657-8501, Japan}}

\newcommand{\kit}{\affiliation{Institute for Astroparticle Physics, Karlsruhe Institute of Technology, 76021 Karlsruhe, Germany}}
\newcommand{\tsinghua}{\affiliation{Department of Physics \& Center for High Energy Physics, Tsinghua University, Beijing 100084, China}}

\begin{document}

\title{Double-Weak Decays of $^{124}$Xe and $^{136}$Xe in the XENON1T and XENONnT Experiments}

\author{E.~Aprile}\columbia
\author{K.~Abe}\tokyo
\author{F.~Agostini}\bologna
\author{S.~Ahmed Maouloud}\paris
\author{M.~Alfonsi}\mainz
\author{L.~Althueser}\munster
\author{B.~Andrieu}\paris
\author{E.~Angelino}\torino
\author{J.~R.~Angevaare}\nikhef
\author{V.~C.~Antochi}\stockholm
% ...

\collaboration{XENON Collaboration}
\email[]{xenon@lngs.infn.it}
\noaffiliation

\date{\today}

\begin{abstract}
\newpage
\noindent We present results on the search for two-neutrino double-electron capture ($\dec$) of \isotope[124]{Xe} and neutrinoless double-$\upbeta$ decay ($\doublebeta$) of \isotope[136]{Xe} in XENON1T.
We consider captures from the K- up to the N-shell in the $\dec$ signal model  and measure a total half-life of $T_{1/2}^{\dec}=(1.1\pm0.2_\text{stat}\pm0.1_\text{sys})\times 10^{22}\;\text{yr}$ with a $0.87\;\text{kg}\times\text{yr}$ isotope exposure.
The statistical significance of the signal is $7.0\,\sigma$.
We use XENON1T data with $36.16\;\text{kg}\times\text{yr}$ of \isotope[136]{Xe} exposure to search for $\doublebeta$.
We find no evidence of a signal and set a lower limit on the half-life of $T_{1/2}^{\doublebeta} > 1.2 \times 10^{24}\;\text{yr}\; \text{at}\; 90\,\%\;\text{CL}$. This is the best result from a dark matter detector without an enriched target to date.
We also report projections on the sensitivity of XENONnT to $\doublebeta$.
Assuming a  $275\;\text{kg}\times\text{yr}$ $^{136}$Xe exposure, the expected sensitivity is $T_{1/2}^{\doublebeta} > 2.1 \times 10^{25}\;\text{yr}\; \text{at}\; 90\,\%\;\text{CL}$, corresponding to an effective Majorana mass range of $\langle m_{\upbeta\upbeta} \rangle  < (0.19 - 0.59)$\;eV/c$^2$.
\end{abstract}

\graphicspath{{xenon1t_0vbb/plots/}{xenon1t_dec/plots/}{xenonnt_0vbb/plots/}}

\maketitle

\section{Introduction}
\label{sec:intro}

The XENON collaboration acquired science data with the XENON1T experiment at the INFN Laboratori Nazionali del Gran Sasso (LNGS) in Italy from November 2016 until December 2018.
Its primary goal was the search for interactions between xenon nuclei and dark matter (DM) in the form of weakly interacting massive particles (WIMPs)~\cite{aprile2017xenon1t,xenon1t:sr1:wimp}. In addition to these nuclear recoils, the detector was also sensitive to other rare processes that could be measured as energy depositions on atomic electrons in xenon, electronic recoils (ER).
In particular, the collaboration reported the first direct observation of the two-neutrino double-electron capture ($2 \upnu\mathrm{ECEC}$) in $^{124}$Xe with $4.4\,\upsigma$ significance~\cite{XENON:2019dti}.
The low background rate of the experiment and its good energy reconstruction and resolution up to the MeV region~\cite{Aprile:2020yad} also allow for the search for the neutrinoless double-$\upbeta$ decay ($\doublebeta$) of \isotope[136]{Xe}.
This potential will be extended with XENONnT, the latest experiment within the XENON program, owing to its approximately three times larger active xenon mass and six times smaller background rates \cite{xenonnt_mc:Aprile_2020}.

The yet unobserved $\doublebeta$ is a nuclear transition predicted by extensions of the Standard Model (SM).
Two neutrino double-$\upbeta$ decay ($\twonubeta$) is allowed in the SM and has been observed in \isotope[136]{Xe} making it a candidate isotope to search for a $\doublebeta$ peak at the $Q$-value of $Q_{\upbeta\upbeta}=(2457.83 \pm 0.37)$\,keV~\cite{PhysRevLett.98.053003,q-value:McCowan:2010zz}.
The currently best lower limit on the \isotope[136]{Xe} $\doublebeta{}$ half-life, $T^{0\upnu}_{1/2}$ is set by KamLAND-Zen, $T^{\doublebeta}_{1/2} > 2.3 \times 10^{26}$\,yr~\cite{KamLAND-Zen:2022tow} at 90\% confidence level (CL). 

A detection of $\doublebeta{}$ or neutrinoless double-electron capture ($0\upnu\mathrm{ECEC}$) would demonstrate the violation of total lepton number and prove the existence of a non-zero Majorana component of neutrino mass.
Under the assumption of light Majorana neutrino exchange, the half-life is related to the effective Majorana mass, $\langle m_{\upbeta\upbeta}\rangle$, by~\cite{Dolinski:2019nrj}
\begin{equation}
    \langle m_{\upbeta\upbeta}\rangle ^2 = \frac{m_\text{e}^2}{G_{0\upnu}\left|M_{0\upnu} \right|^2T_{1/2}^{0\upnu}}.
    \label{eq:0vbb:effective_mass}
\end{equation}
Here, $G_{0\upnu}$ is the phase-space factor in units of $\text{yr}^{-1}$~\cite{PhysRevC.85.034316}, $M_{0\upnu}$ is the dimensionless nuclear matrix element (NME), and $m_\text{e}$ is the electron mass in $\text{eV}/\text{c}^2$.
Since $\langle m_{\upbeta\upbeta}\rangle$ can contain phase cancellations from the Pontecorvo-Maki-Nakagawa-Sakata (PMNS) matrix, it is sensitive to the neutrino mass hierarchy~\cite{Esteban:2020cvm,nufit50}.
While the phase-space factor can be calculated with relative precision, theoretical uncertainties are associated with the choice of the NME.
The central values of the most extreme \isotope[136]{Xe} NMEs presented in~\cite{Engel:2016xgb} range from ${M_{0\upnu}=1.550 \text{–} 4.773}$~\cite{Mustonen:2013zu, LopezVaquero:2013yji} and are considered when interpreting $\doublebeta{}$ decay limits in this work.
This illustrates that $M_{0\upnu}$ is a major source of uncertainty on $\langle m_{\upbeta\upbeta}\rangle$. Although there is no direct correspondence between the neutrinoless and two-neutrino NMEs, the measured half-lives of two-neutrino decays such as $\dec$ can be used as a benchmark for different NME calculation approaches~\cite{Engel:2016xgb}.

In this work, we perform $2\upnu\text{ECEC}$ and $\doublebeta{}$ peak searches in the measured ER energy spectrum of XENON1T and assess XENONnT's sensitivity to $\doublebeta{}$ of \isotope[136]{Xe} using simulated data. The paper is organized as follows.
Sec.~\ref{sec:common_info} and Sec.~\ref{subsec:detector} give an overview of the XENON1T detector and the XENONnT detector, respectively.
Sec.~\ref{mc:common} highlights the background components relevant for the 2$ \upnu\mathrm{ECEC}$ and $\doublebeta{}$ decay searches and their constraints for background simulations.
Sec.~\ref{fits:common} details the fitting method which is employed to derive results.
Sec.~\ref{sec:dec} summarizes an updated search for 2$ \upnu\mathrm{ECEC}$ in \isotope[124]{Xe} following an extension of the signal model to include captures from higher electron orbitals and using a larger exposure compared to the previous analysis~\cite{XENON:2019dti}.
Sec.~\ref{sec:1t_0vbb} reports on a $^{136}$Xe $\doublebeta{}$ decay search in XENON1T.
Sensitivity projections for the XENONnT experiment are discussed in Sec.~\ref{sec:nt_0vbb}.
A summary of the results and an outlook on $\doublebeta{}$ search with WIMP detectors are given in Sec.~\ref{sec:conclusion}.

\section{General aspects of the analyses}
XENON1T~\cite{aprile2017xenon1t} and XENONnT are designed as dual-phase xenon time projection chambers (TPC). These cylindrical detectors are filled with liquid xenon (LXe) and have a thin xenon gas layer at the top. Several electrodes in the liquid and gas enable the application of electric fields. Photomultiplier tubes (PMTs) at the top and bottom read out the signals from particle interactions.
A particle interaction in LXe produces excitation, ionization and heat.
The total number of measurable quanta depends on the energy deposition of the incident particle and the interaction type, e.g., nuclear recoil or electronic recoil.
Excitation occurs in the form of excited xenon dimers that decay to the ground state by emitting scintillation light at 175~nm~\cite{175nm:FUJII2015293}.
The electron-ion pairs from the ionization process can recombine leading to further light emission.
The resulting primary scintillation signal is registered by the PMTs and denoted S1.
Full recombination is suppressed by an electric drift field that moves the electrons away from the interaction site and towards the liquid-gas interface.
There they are accelerated into the gas gap by an extraction field and produce a secondary scintillation signal, S2, proportional to the number of extracted electrons. In these analyses, we consider events from particle interactions that have a summed PMT waveform containing at least one S1 and S2 pair. The time scale for such events is given by the maximum drift time of $\mathcal{O}(1\;\text{ms})$ in both detectors.

The 3D-position of an interaction is reconstructed using the distribution of the S2 light in the top PMT array ($x,\,y$) and the time delay between the S1 and S2 signals ($z$). Using the self-shielding of xenon and this position information allows for the definition of a fiducial volume (FV) with reduced background levels from external sources, i.e., located outside the detector and the detector materials themselves.
In XENON1T data, S1 and S2 signal sizes were corrected accounting for the position-dependent measurement efficiencies.
Moreover, the measured interaction positions were corrected for inhomogeneities of the drift field~\cite{Aprile:2019bbb}.

The total deposited energy of an event is characterized by the weighted sum of corrected S1 and corrected S2.
The energy calibration in XENON1T was performed using monoenergetic peaks.
The anti-correlation between the S1 and S2 signals at different energies allowed us to compute the photon detection efficiency $g_1$ and the charge amplification factor $g_2$~\cite{Aprile:2020yad}.
These parameters were used as weighting factors for the two observables S1 and S2 in the energy calibration, which we call combined energy scale (CES). The analyses in the following sections were carried out in the CES parameter space.

\subsection{The XENON1T experiment}
\label{sec:common_info}
The XENON1T~\cite{aprile2017xenon1t} TPC had a height of 97\;cm and a diameter of 96\;cm.
Two arrays of 127 and 121 Hamamatsu R11410-21 3-inch PMTs were arranged above and below the sensitive volume of the TPC, respectively.
The active volume consisted of 2\;t of LXe out of a total of 3.2\;t in the detector.
The TPC side walls were made of Polytetrafluoroethylene (PTFE) reflective panels to enhance the light collection efficiency. Two electrodes, a cathode placed at the bottom of the TPC and a gate $\sim$2.5\;mm below the liquid-gas interface, produced a drift field of 81\;V/cm.
An anode placed 2.5\;mm above the liquid-gas interface created an extraction field of 8.1\;kV/cm.
The cryostat was immersed at the center of a stainless-steel tank, filled with 700\;t of ultra-pure demineralized water, used to shield environmental gammas and neutrons. The tank was instrumented with 84 PMTs to actively tag muons and muon-induced backgrounds through the detection of Cherenkov light.

The data used for the analyses presented here were acquired between February 2017 and September 2018 during the main science run of the experiment (SR1) and a second run (SR2) targeting research and development.
Subsets of SR1 and SR2 were selected for the specific analyses and are described in more detail in Sec.~\ref{sec:dec} and~\ref{sec:1t_0vbb}.

\subsection{The XENONnT experiment}
\label{subsec:detector}
XENONnT is the successor experiment to XENON1T. 
It was commissioned in the second half of 2020 and started operations shortly thereafter. 
It reuses several subsystems already developed for XENON1T, with additional radon removal, LXe purification, neutron veto  and xenon gas storage systems.
The TPC has an active region of 133\;cm in diameter and 148\;cm in height containing 5.9\;t of LXe. The cryostat holds 8.4\;t of LXe in total.
Two hexagonal arrays contain 253 and 241 PMTs at the top and bottom, respectively.
In order to avoid digitizer saturation from large S2s at MeV energies, the PMTs in the top array are read out with an amplification factor of $\times0.5$, in parallel to the $\times10$ amplification used for DM searches. This secondary readout was specifically installed for $\doublebeta{}$ searches.
Muons are suppressed by means of the same tagging system developed for XENON1T.
Additionally, a novel neutron veto (NV) system uses 120 PMTs inside an optically-separated volume around the cryostat to detect signals originating from the capture of radiogenic neutrons~\cite{xenonnt_mc:Aprile_2020}. For the projections in section \ref{sec:nt_0vbb}, we assume the same detector operating conditions as in~\cite{xenonnt_mc:Aprile_2020}.

\subsection{Electronic Recoil Background}
\label{mc:common}
The analyses presented in Sec.~\ref{sec:dec},~\ref{sec:1t_0vbb} and~\ref{sec:nt_0vbb} require modeling the individual background sources via Monte Carlo (MC) simulations.
The simulated backgrounds include radioactive impurities in the detector components and the xenon target itself, as well as solar neutrinos~\cite{XENON:2019dti,lowER:Aprile:2020tmw}.
The background composition in each analysis depends on the energy range and the chosen FV.
In the following, we describe simulation aspects and backgrounds that are common to the $\dec$ and $\doublebeta$ analyses. More specific background contributions are discussed in the respective sections.

The background energy spectra were obtained with the XENON1T~\cite{mc:xenon1t:XENON:2015gkh} and XENONnT~\cite{xenonnt_mc:Aprile_2020} MC simulation frameworks, respectively.
First, energy depositions from radioactive decays were simulated using the implementation of the detector geometry in Geant4~\cite{AGOSTINELLI2003250,1610988,ALLISON2016186}.
Next, individual energy depositions from the same Geant4 event were clustered based on their relative S2 sizes and their $z$-separation.
This clustering mimics the finite resolution in the reconstruction of multiple nearby energy depositions in the detector.
Events with a single energy cluster are denoted as single-site (SS) events while those with multiple clusters are multi-site (MS).
Measured waveforms of SS and MS events would contain a single S1 signal but, while SS events have only one S2, MS events may contain multiple resolved S2s from energy depositions at different depths.
The S2-based MS event resolution deteriorates for deeper events due to longitudinal diffusion of the electron cloud. Clustering distances for XENON1T range from 6.5\;mm at the top to 11.5\;mm at the bottom of the TPC. These were determined using simulated waveforms. Events with multiple S2s at the same depth can be identified using PMT hit pattern information. This information was not included in the simulations and accordingly not used for MC event selections.

After the clustering, SS events were selected from the MC simulation within the FV chosen for each analysis.
The resulting energy spectra were convolved with a Gaussian representing the measured energy resolutions of the SR1 and SR2 data~\cite{Aprile:2020yad, lowER:Aprile:2020tmw}.
Monoenergetic peaks from target-intrinsic sources without significant Compton-scattering or Bremsstrahlung contributions were modeled as single Gaussian lines with the standard deviation given by the energy resolution.
 
The material backgrounds considered in XENON1T and XENONnT originate mainly from the \isotope[238]{U} and \isotope[232]{Th} decay chains as well as from \isotope[40]{K} and \isotope[60]{Co}.
Their contributions were constrained using the radioassay results of the XENON1T~\cite{screening:XENON:2017fdb} and XENONnT~\cite{XENON:2021mrg} detector materials.
Due to its 5.27~yr half-life, the decay of \isotope[60]{Co} after the radioassay, is taken into account.
We assume that production of \isotope[60]{Co} by activation underground is negligible.
In the MC simulation of the uranium and thorium decay chains, we take possible decay chain disequilibrium into account.
The early and late parts of the uranium chain were split at \isotope[226]{Ra}, and the \isotope[232]{Th} decay chain was split at \isotope[228]{Th}. For the XENON1T analysis, the full \isotope[232]{Th} chain was simulated. In order to account for disequilibrium, the partial chain starting at \isotope[228]{Th} was also simulated. In fits of simulated background spectra to measured data it could be added or subtracted from the full chain in the fit, depending on the observed disequilibrium. This different treatment with respect to the \isotope[238]{U} chain was caused by the internal processing of the decay chains and ensured that all expected $\upgamma$-rays were present in the simulations. For XENONnT, both parts of the \isotope[232]{Th} chain were treated independently.

After applying the FV selection, no further spatial information was included in the background model.
The energy distributions from the same isotope but originating from different materials are essentially identical.
Therefore, the relative contributions from different materials to the background from each isotope were fixed in the analysis using the screening measurements.
With this, a single scaling parameter for each isotope was required in each background model.

The \isotope[222]{Rn} emanation from materials of the detector to the LXe target induces an intrinsic background contribution. The two most relevant radon daughter isotopes identified for the analyses presented in this work are \isotope[214]{Pb} and \isotope[214]{Bi}.
The latter predominantly undergoes $\upbeta$-decay to \isotope[214]{Po}~\cite{a214:WU2009681}.
Subsequently, \isotope[214]{Po} decays via $\upalpha$-emission with a half-life of 164~$\upmu$s. The close timing coincidence of the two decays, with respect to the event time scale of $\mathcal{O}(1\;\text{ms})$,  allows for their tagging and enables effective rejection of \isotope[214]{BiPo} events inside the active volume. A rejection efficiency of 99.8~$\%$ is assumed for XENONnT. For XENON1T, the \isotope[214]{Bi} background from radon emanation is discussed in section~\ref{sec:bkg_1t}.

The $\twonubeta$ decay of \isotope[136]{Xe} features a continuous energy spectrum with the endpoint at $Q_{\upbeta\upbeta}$.
Theoretically calculated distributions of the energies and relative emission angles of the two electrons~\cite{PhysRevC.85.034316,dec-tretyak} were used as input to Geant4. The IBM-2 higher-state dominance (HSD)~\cite{PhysRevC.85.034316} model of the $\twonubeta{}$ process was used. The resulting energy spectrum was normalized according to the expected decay rate corresponding to a half-life of $T_{1/2}^{2\upnu\upbeta\upbeta} = (2.165 \pm 0.016_\text{stat} \pm 0.056_\text{sys})\times 10^{21}$\;yr~\cite{PhysRevC.89.015502} and the measured isotopic abundance $\eta_{^{136}\text{Xe}}=(8.49\pm0.04_\text{stat}\pm0.13_\text{sys})\times10^{-2}\;\text{mol}/\text{mol}$ in XENON1T.
Considering these uncertainties, the background contribution over the whole energy range can be constrained with a relative uncertainty of $3\,\%$. The difference between the HSD spectrum and an alternative single-state dominance (SSD) spectrum~\cite{private:Kotila} was not considered as a source of systematic uncertainty due to the subdominant contribution of this background in the analyses.
Due to a combination of sub-percent energy resolution at $Q_{\upbeta\upbeta}$~\cite{Aprile:2020yad} and low decay rate, the $\twonubeta{}$ contribution in the region of interest (ROI) is expected to be several orders of magnitude lower than the material background for both XENON1T and XENONnT.
\subsection{Fit method and limit setting}
\label{fits:common}

The results presented below were derived using the standard procedure of a Poisson binned log-likelihood where nuisance parameters are profiled~\cite{Cowan:2010js}.
The likelihood reads
\begin{align}
    \mathcal{L}\left(\mu_{s}, \vec{\theta}\right) = \prod_{{i}}^{\text{bins}}\text{Poisson}\left(N_{\text{i}},\,\lambda_{\text{i}}(\vec \theta ) + n^s_{\text{i}}(\mu_s, \vec{\theta}) \right) \times \nonumber\\  \prod_{\text{j}}^\text{constraints} \text{Gauss}\left(\theta_{\text{j}},\,\mu_{\text{j}},\,\sigma_{\text{j}}\right).
    \label{eq:0vbb:poissonlikelihood}
\end{align}
$N_{\text{i}}$ is the measured number of events in each energy bin, $\lambda_{\text{i}}$ is the number of expected background events as a function of the nuisance parameters $\vec{\theta}$.
The number of expected signal events in bin $i$ is denoted with $n_i^s$ and depends on the signal strength $\mu_s$ and the nuisance parameters $\vec{\theta}$.
Each constrained nuisance parameter $\theta_j$ has an expected mean value $\mu_\text{j}$ with a standard deviation $\sigma_\text{j}$. 
Details regarding the set of nuisance parameters for each analysis are given in the specific sections. The binned likelihood for signal and background model fits to measured data in Secs.~\ref{sec:dec} and~\ref{sec:1t_0vbb} are normalized such that it can be interpreted as in a $\chi^2$ fit~\cite{chi2lambda:BAKER1984437}, e.g., in terms of goodness of fit.
Following the nomenclature in \cite{chi2lambda:BAKER1984437}, this goodness of fit measure is labeled $\chi^2_\lambda$. 

The test statistic for the \doublebeta{} searches in Secs.~\ref{sec:1t_0vbb} and~\ref{sec:nt_0vbb} is
\begin{equation}
\label{eq:test_statistics}
    q\left(\mu_{s}\right)=-2 \ln \frac{\mathcal{L}\left(\mu_{s}, \hat{\hat{{\vec{\theta}}}}\right)}{\mathcal{L}\left(\hat{\mu}_{s}, \hat{\vec{\theta}}\right)},
\end{equation}
where quantities with a single hat denote the set of parameters which correspond to the unconditional maximum of the likelihood while quantities with two hats denote the set of parameters maximizing the conditional likelihood.
Under certain conditions the test statistic $q(\mu_s)$ follows an asymptotic distribution which is given by a $\chi^2$ distribution with one degree of freedom~\cite{Cowan:2010js}. The distribution of $q(\mu_s)$ is estimated by toy-MC simulations to validate the assumption of asymptoticity.
We report only the upper edge of the Feldman-Cousins confidence interval if an excess is smaller than 3\,$\sigma$.
Similar to \cite{xenon1t:analysis:2019izt} this imposes overcoverage for very small signals.
The $3\,\sigma$ significance threshold only serves as the transition point between reporting one- and two-sided intervals, and was decided prior to the analysis to ensure correct coverage.
All derived limits correspond to a $90\,\%\;\text{CL}$.

The parameter of interest is the event rate $A_{\upbeta\upbeta}$ of the double-weak processes $\dec$ or $\doublebeta$. For a measured $A_{\upbeta\upbeta}$, the half-life is
\begin{equation}
    T_{1/2}^{\upbeta\upbeta} = \ln{2}\times\frac{N_A \times \eta_{\text{Xe}} \times \epsilon_\text{SS} }{{A_{\upbeta\upbeta}}\times M_A},
    \label{eq:0vbb:halflife:calc}
\end{equation}
where $M_A=0.131\;\nicefrac{\text{kg}}{\text{mol}}$ is the xenon molar mass, $N_A$ is Avogadro's constant, and $\eta_{\text{Xe}}$ is the isotopic abundance of the xenon isotope, \isotope[124]{Xe} or \isotope[136]{Xe}, under consideration. The SS efficiency $\epsilon_\text{SS}$ is the fraction of signal events that are identified as SS events. With an electron mean free path in LXe smaller than 3\;mm, the majority of the two electrons emitted in the \doublebeta{} decay are detected as SS events. For reference, the spatial ($z$) resolution of XENON1T is $\sim8\;\text{mm}$ for S2 signals larger than 10$^3$ photo-electrons (PE)~\cite{Aprile:2019bbb}. An efficiency loss occurs when Bremsstrahlung is emitted by one of the double-$\upbeta$ electrons leading to an MS event.
The SS efficiency for $0\upnu\upbeta\upbeta$ was estimated with MC to be $90.3\,\%$ in XENON1T and $91.0\,\%$ in XENONnT.
For the X-rays and Auger electrons emitted in the double-electron capture, the SS efficiency is 100\,\% due to the sub-millimeter mean free path of the quanta.

\section{Extended search for $^{124}$Xe two-neutrino double-electron capture in XENON1T}
\label{sec:dec}
\begin{figure}[t]
  \centering
  \includegraphics[width=\linewidth]{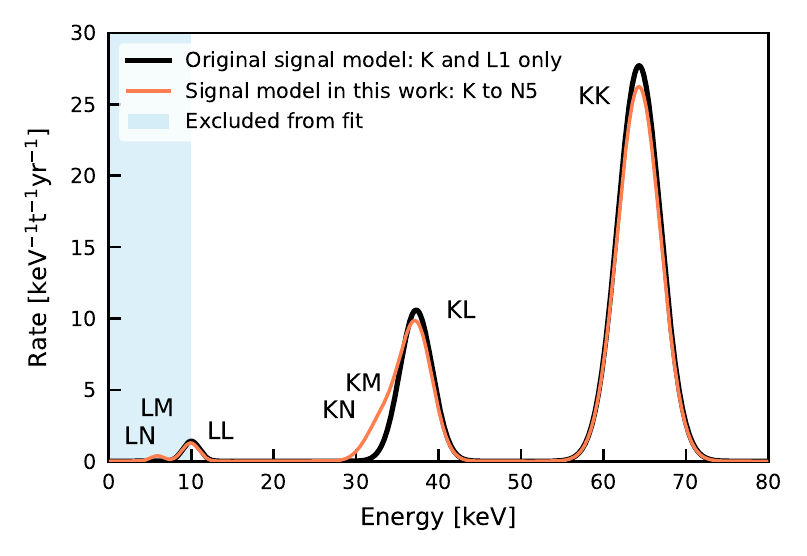}
  \caption{Comparison of the two $\dec$ signal models discussed in this work: the model from~\cite{lowER:Aprile:2020tmw} considering K- and L1-shells only (black) and the updated model considering shells from K to N5 (orange). The peak widths reflect the energy resolution of XENON1T. Uncertainties associated with the peak scaling, the peak positions and the peak widths are not shown for visibility. The energy region that was excluded due to the low-energy excess reported in~\cite{lowER:Aprile:2020tmw} is marked in blue.}
  \label{fig:edec:signal_comparison}
\end{figure}

\begin{table}[b]  
\centering
  \begin{center}
  \caption{Relative capture fractions and energies of double-electron captures from different shells. The capture fractions considering only the K- and L1-shells were used in~\cite{XENON:2019dti}. These were obtained using the Dirac solutions of the bound electron wave functions for a finite size nucleus in table V of~\cite{dec-Doi:1991xf} as in~\cite{xmass:Abe:2018gyq, lowER:Aprile:2020tmw}. The improved double-capture fraction calculation in this work considers all shells up to the N5-shell with the squared amplitudes of the radial wave functions in~\cite{toi:firestone_shirley_1999}. The individual subshells are added together in the table rows. The energy ranges for the captures are from the orbital with the lowest to that with the highest binding energy. The captures with energies below the analysis energy threshold of 10~keV, namely LM-, LN-, MM-, MN- and NN-capture, are subsumed under the label other. 
  }
      \label{tab:edec:fractions}
      {
      \tabcolsep=1.00mm
      \begin{ruledtabular}
      \begin{tabular}{l c c c}
      Decay & K \& L1\;[\%] & K to N5\;[\%] & Energy\;[keV]\\
      \colrule
    KK & $76.5$ & $72.4$ & $64.3$ \\
    KL$_{1,2,3}$ & $22.0$ & $20.0$ & $36.7-37.3$ \\
    KM$_{1,2,3,4,5}$ & $-$ & $4.3$ & $32.9-33.3$ \\
    KN$_{1,2,3,4,5}$ & $-$ & $1.0$ & $32.3-32.4$ \\
    L$_{1,2,3}$L$_{1,2,3}$ & $1.6$ & $1.4$ & $8.8-10.0$ \\
    Other & $-$ & $0.8$ & $<10$ \\
      \end{tabular}
      \end{ruledtabular}}
  \end{center}
\end{table}
The $\dec$ analysis presented here builds on the previous result~\cite{XENON:2019dti} with an increased exposure and additionally considers double-electron capture contributions from higher atomic shells. As in single-electron capture, double-electron capture rates chiefly depend on the overlap between the electron and nuclear wave functions~\cite{electron:capture:RevModPhys.49.77}. Since the electrons in the s-orbitals of the K- and L-shell (L1) feature the largest overlap, only they have been commonly considered in theoretical studies and in the interpretation of experimental data~\cite{dec-Doi:1991xf, Suhonen:2013rca, lowER:Aprile:2020tmw, xenonnt_mc:Aprile_2020}.
However, with xenon's 54 atomic electrons, the \mbox{M-,} N- and O-shells with s-, p-, d- and higher orbitals should also contribute to the total double-electron capture decay rate. Values for the squared amplitudes of the radial wave functions up to the N5-shell are tabulated in~\cite{toi:firestone_shirley_1999}. The corresponding relative capture fractions are given in Tab.~\ref{tab:edec:fractions} together with the results when considering K- and L1-captures only. The respective signal models are illustrated in Fig.~\ref{fig:edec:signal_comparison}. Taking the additional shells into account slightly decreased the relative fraction of KK-, KL- and LL-captures with respect to all decays.
No literature values for the $O_{1,2,3}$ wave functions were available. Their values were approximated by scaling the $N_{1,2,3}$ values with a factor 1/4 -- approximately the scaling between the tabulated M- and N-shell wave functions. With this, O-captures were estimated to present $\mathcal{O}(0.1)\,\%$ corrections to the other capture fractions. Accordingly, we do not include O-captures in the signal model.
We note that this approach is still simplified compared to the calculation approach for single-electron capture outlined in~\cite{electron:capture:RevModPhys.49.77}, where the Q-value of the decay as well as the energies, parities, and angular momenta of the nuclear states involved in the decay are considered.
However, more work is needed to extend this treatment for double-electron capture. The simplified treatment was considered as a systematic uncertainty in the analysis. 
\begin{table}[t]  
\centering
  \begin{center}
  \caption{Overview of the four datasets considered in this analysis. Each FV mass uncertainty contains both the analytic mass calculation uncertainty from the LXe density~\cite{nist:webbook:https://doi.org/10.18434/t4d303} and the detector dimensions, as well as the systematic uncertainty from the position reconstruction. The mass value in the table represents the mean of the analytically calculated mass and the mass determined from \isotope[83\text{m}]{Kr} calibration data. 
 }
      \label{tab:edec:datasets}
      {
      \tabcolsep=1.00mm
      \begin{ruledtabular}
      \begin{tabular}{l c c c}
      Science run & Live time\;[d] & Dataset & FV mass\;[kg]\\
      \colrule
    SR1$_\text{a}$ & 171.2 & SR1$_\text{a}^\text{in}$ & $1032 \pm 4$ \\
     & & SR1$_\text{a}^\text{out}$ & $460\pm4$ \\
    SR1$_\text{b}$ & 55.8 & SR1$_\text{b}^\text{in}$ & $1029 \pm$ 3\\
    SR2 & 24.3 & SR2 & $1033 \pm 5$ \\
      \end{tabular}
      \end{ruledtabular}}
  \end{center}
\end{table}

\subsection{Reconstruction and cuts}~\label{sec:edec:cuts}
The data used for this analysis comprises 226.9 live-days from SR1, subdivided into two partitions with nominal (SR1$_\text{a}$) and increased (SR1$_\text{b}$) background produced by neutron calibrations as in~\cite{lowER:Aprile:2020tmw}.
Moreover, 24.3~d from SR2 were added to the dataset. The total exposure of the combined dataset is $0.93\;\text{t}\times\text{yr}$ of which $0.68\;\text{t}\times\text{yr}$ in SR1$_\text{a}$ overlap with the data used in~\cite{XENON:2019dti}. With a measured abundance of $\eta_{^{124}\text{Xe}} = (9.94\pm0.14_\text{stat}\pm0.15_\text{sys})\times10^{-4}\;\text{mol}/\text{mol}$ the \isotope[124]{Xe} isotope exposure is $0.87\;\text{kg}\times\text{yr}$.
The energy calibration parameters from~\cite{Aprile:2020yad} were used for the SR1 data after an upgrade of the XENON1T data processor.
They were derived anew for SR2 following the same method as laid out in~\cite{Aprile:2020yad}.
Event positions were reconstructed with a neural network. 
As in~\cite{XENON:2019dti}, the SR1$_\text{a}$ dataset was analyzed in a 1.5~t superellipsoid FV which was subdivided into an inner 1.0~t cylinder and an outer 0.5~t shell.
The SR1$_\text{b}$ and SR2 datasets were analyzed in cylindrical 1\;t FVs~\cite{aprile2017xenon1t} only. For SR2 no outer FV was used, since the respective cuts and energy calibration had been defined for the 1~t cylinder. For SR1$_\text{b}$ the addition of an outer FV was not expected to yield a significant increase in sensitivity due to the larger background from neutron calibrations. The reconstruction-induced systematic uncertainties on the masses for all FVs were estimated using homogeneously distributed \isotope[83\text{m}]{Kr} calibration data and are below $1\,\%$. An overview of the four datasets with their respective live times and FV masses is given in Tab.~\ref{tab:edec:datasets}.

The data quality criteria (cuts) from~\cite{lowER:Aprile:2020tmw} were applied to the SR1 data. Two cuts using PMT hit pattern information were adapted for SR2, taking into account changes in the PMT configuration between SR1 and SR2. A cut evaluating the difference in reconstructed event positions from two different algorithms was discarded since not all position reconstruction algorithms had been updated for SR2. As the cut targets singular outlier events, its omission has no impact.
The energy ranges employed for the analysis are 10–200~keV for the central 1\;t FV and 10–160 keV for the outer 0.5\;t volume. The lower fit bounds were chosen in order to exclude the \mbox{low-energy} excess observed in~\cite{lowER:Aprile:2020tmw}. Above 160 keV and 200 keV in the inner and outer FVs, respectively, Compton scattering lead to MS events. We were therefore unable to obtain a clean calibration sample to determine the acceptance of our SS cuts. Thus, we set these energies as our upper fit bounds.

In~\cite{XENON:2019dti}, misidentified events from the intrinsic calibration isotope~\isotope[83\text{m}]{Kr} led to the presence of a secondary lower-energy peak from this isotope (cf. Sec.~\ref{sec:edec:bgmodel}). This obscured possible KL-, KM- and KN-capture peaks. A new cut from~\cite{lowER:Aprile:2020tmw} reduces these events to the $10^{-4}$ level. This enabled the addition of the double-electron capture peaks from higher shells to the signal model.

\subsection{$\mathbf{^{125}}$I background}\label{sec:edec:i125}
 A key background in this search comes from \isotope[125]{I}, a daughter isotope of \isotope[125]{Xe} which is produced by neutron capture on \isotope[124]{Xe} in the detector as well as in the detector-external xenon purification loop~\cite{xmass:Abe:2018gyq, XENON:2019dti}. The half-life of \isotope[125]{Xe} is 16.9~h~\cite{a125:KATAKURA2011495} and its electron capture decay with subsequent $\upgamma$-emission was observed outside of the energy region used for the $2\upnu\text{ECEC}$ search. 
The half-life of $\isotope[125]{I}$ is 59.4~d and its three atomic deexcitation cascades from K-, L- and M-shell electron captures are merged with the 35.5~keV $\upgamma$-ray from the daughter \isotope[125]{Te}~\cite{a125:KATAKURA2011495, i125k:ToRad} resulting in a peak energetically close to the $2\upnu\text{ECEC}$ signal peaks. The relative fractions for K-, L-, and M-shell electron captures as well as the merged peak energies are given in Tab.~\ref{tab:edec:i125}.

The absolute \isotope[125]{I} background contribution in SR1$_\text{a}$ and SR1$_\text{b}$ was constrained by integrating the activity model from~\cite{XENON:2019dti} over the new data selection.
The integration yielded $N^{^{125}\text{I}}_{\text{SR1}_\text{a}}=(10\pm5)\;\text{t}^{-1}$ and $N^{^{125}\text{I}}_{\text{SR1}_\text{b}}=(100\pm20)\;\text{t}^{-1}$.
The SR2 model was derived with the same method as in~\cite{XENON:2019dti}, which tracks the time evolution of the \isotope[125]{Xe} parent activity. The number of \isotope[125]{I} nuclei $N_{^{125}\text{I}}$ in the TPC is connected to the number of \isotope[125]{Xe} nuclei $N_{^{125}\text{Xe}}$ by the differential equation
\begin{align}
    \frac{\text{d}N_{^{125}\text{I}}}{\text{d}t} &= \lambda_{^{125}\text{Xe}}N_{^{125}\text{Xe}}(t) -  \left(\lambda_{^{125}\text{I}}+\frac{1}{\tau_\text{pur}}\right)N_{^{125}\text{I}}(t)\nonumber\\
    &=\lambda_{^{125}\text{Xe}}N_{^{125}\text{Xe}}(t) -  \frac{1}{\tau_\text{eff}}\,N_{^{125}\text{I}}(t).
\end{align}
The decay constants $\lambda_{^{125}\text{Xe}}$ and $\lambda_{^{125}\text{I}}$ describe the production and decay of \isotope[125]{I}, respectively. The purification time constant $\tau_{\text{pur}}$ accounts for continuous iodine removal in the xenon purification loop and is a fit parameter in the model. Combining decay and purification leads to an effective time constant $\tau_\text{eff}$.

\begin{figure}[tp]
  \centering
  \includegraphics[width=\linewidth]{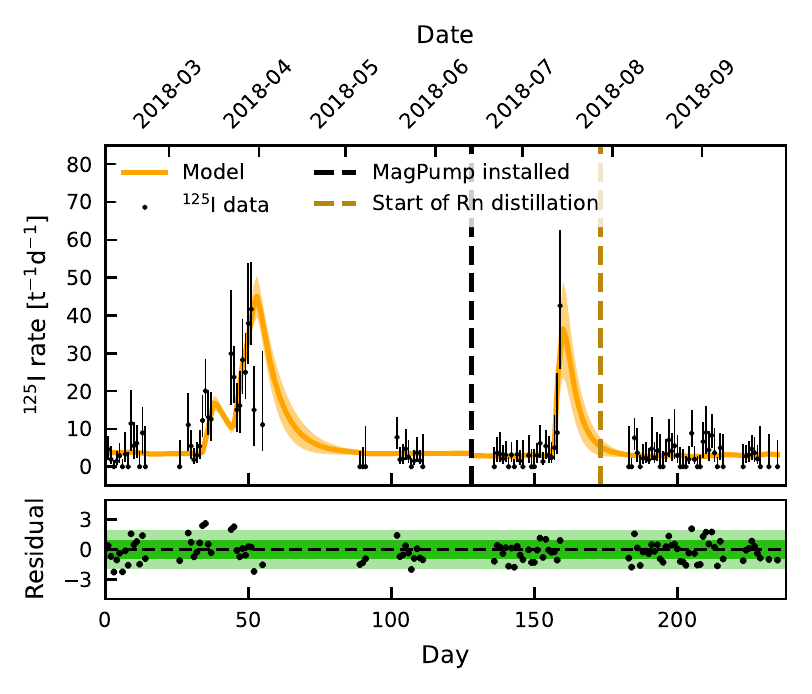}
  \caption{Fit of the SR2 \isotope[125]{I} model to data with $\chi^2_\lambda/\text{ndf}=138/108$.
  The data from a $2\,\sigma_E$ interval around the \isotope[125]{I} peak at 67.3~keV is subdivided into 1-day bins shown by the black markers.
  The rate was corrected for statistical coverage and live-time. 
  The $1\,\sigma$ model uncertainty is shown as an orange band around the solid orange best-fit line.
  The peaks in the rate are caused by three neutron generator calibration campaigns producing \isotope[125]{I}.
  The installation of the magnetically-coupled piston pump (MagPump)~\cite{radondistillation:XENON100:2017gsw}, indicated by the dashed black line, marks the separation point of the two model periods. 
  The start of the radon distillation 
  is shown by the dashed golden line.}
  \label{fig:edec:i125}
\end{figure}

The replacement of the purification system's pumps with a new ultra-clean magnetically-coupled piston pump~\cite{pump:Brown:2018uya} during SR2 allowed increasing the purification flow from $(50\pm2)\;\text{slpm}$ to $(79\pm2)\;\text{slpm}$.
As the iodine removal was expected to be proportional to the purification flow, the ratio of purification time constants $\tau_\text{pur}^\text{SR2,\,early}$ and $\tau_\text{pur}^\text{SR2,\,late}$ could be expressed by the ratio of the purification flows,
 \begin{align}
   r_\text{flow}=\frac{\tau_\text{pur}^\text{SR2,\,late}}{\tau_\text{pur}^\text{SR2,\,early}}=1.59\pm0.08 \text{.}
 \end{align}
With this, the \isotope[125]{I} model for SR2 was divided into a period before and after the pump installation. The pump replacement increased the \isotope[85]{Kr} background level in all subsequent data, most likely due to introduction of airborne krypton to the system during the operation. However, the pump replacement also reduced the \isotope[222]{Rn} background level due to its lower radon emanation~\cite{pump:Brown:2018uya}.
A further reduction was achieved in the final SR2 data when the krypton distillation column was operated in a specialized radon distillation mode~\cite{XENON:2021fkt,radondistillation:XENON100:2017gsw}, but no significant change in the total background rate was observed due to the elevated \isotope[85]{Kr} background.
Consequently, both of these periods were modeled by one background rate parameter. The respective background levels are discussed in the next section. 
The fit of the \isotope[125]{I} data is shown in Fig.~\ref{fig:edec:i125}. The $\isotope[125]{I}$ rate was determined in a 61.4--73.2~keV energy interval around the 67.3~keV K-capture peak, corresponding to twice the energy resolution, $\sigma_E$. The event rate was normalized by the live-time per time bin and the statistical coverage of the $2\,\sigma_E$ interval. Three calibrations with the neutron generator are present in the SR2 data. The three corresponding peaks in rate are well-described by the model. 
The best-fit background in the second model period is reduced by $19.4\,\%$ from $(3.6\pm0.5)\;{\text{t}^{-1}\text{d}^{-1}}$ to $(2.9\pm0.3)\;{\text{t}^{-1}\text{d}^{-1}}$, in accordance with the altered Kr and Rn background levels. The purification time constant $\tau_\text{pur,\,1} = (7\pm3)\;\text{d}$ is compatible with the 7.5\;d turnaround interval of the xenon inventory. Taking into account \isotope[125]{I} decay, the effective time constant $\tau_\text{eff,\,1} = (6.7\pm2.3)\;\text{d}$ agrees with the value of $\tau_\text{eff} = (9.1\pm2.6)\;\text{d}$ in~\cite{XENON:2019dti} within the uncertainties. The constrained flow ratio is reproduced by the fit with $r_\text{flow}=1.59\pm0.07$ and leads to $\tau_\text{pur,\,2} = (5\pm2)\;\text{d}$. The model yields a background expectation of $N^{^{125}\text{I}}_{\text{SR2}}=(3\pm2)\;\text{t}^{-1}$.
\begin{table}[tp]  
\centering
  \begin{center}
      \caption{Relative fractions and energies of \isotope[125]{I} electron capture background peaks. The capture fractions and $\upgamma$-energy were taken from~\cite{i125k:ToRad}. The peak positions are obtained by adding the 35.5~keV $\upgamma$-energy to the energies of the atomic relaxations~\cite{nist-RevModPhys.75.35, xray:LBL}. Uncertainties on the X-ray transition energies are at the eV-level, far below the XENON1T energy resolution and not listed.}
      \label{tab:edec:i125}
      {
      \tabcolsep=1.00mm
      \begin{ruledtabular}
      \begin{tabular}{l c c}
      Decay & Capture fraction\;[\%] & Energy\;[keV]\\
      \colrule
    K+$\upgamma$ & $80.11 \pm 0.17$ & $67.3$ \\
    L+$\upgamma$ & $15.61 \pm 0.13$ & $40.4$ \\
    M+$\upgamma$ & $3.49 \pm 0.07$ & $36.5$ \\
      \end{tabular}
      \end{ruledtabular}}
  \end{center}
\end{table}

\subsection{Additional background sources}\label{sec:edec:bgmodel}
Apart from \isotope[125]{I}, the target-intrinsic isotope \isotope[214]{Pb} from continuous \isotope[222]{Rn} emanation was the dominant background source in this analysis. For its direct $\upbeta$-decay to the \isotope[214]{Bi} ground-state we assume a branching ratio of ${(11.0\pm1.0)}$\,\%~\cite{a214:WU2009681} in our simulations. An approximate activity concentration of \isotope[214]{Pb} was inferred from the $\upalpha$-decay rates of other radon daughters in the detector.
For SR1$_\text{a}$ and SR1$_\text{b}$, the corresponding \isotope[222]{Rn} activity concentration is ${(13.3\pm0.5)\;\upmu\text{Bq}/\text{kg}}$~\cite{xenon1t:radon:2020fbs}.
Due to online radon distillation and the lower emanation of the new pump, the time-averaged \isotope[222]{Rn} activity concentration in SR2 was $(10.1\pm0.3)\;\upmu\text{Bq}/\text{kg}$, $(76\pm4)\,\%$ of the SR1 mean, leading to a reduction in \isotope[214]{Pb} background~\cite{xenon1t:radon:2020fbs}. Since there is no direct measurement of the reduction, the scaling parameter for \isotope[214]{Pb} was left unconstrained in the fit. 
More details can be found in Sec~\ref{sec:bkg_1t}.

Anthropogenic \isotope[85]{Kr} was removed from the XENON1T target by cryogenic distillation before SR1~\cite{XENON:2021fkt}. The natural krypton concentration in xenon was monitored over time by taking regular samples that were measured with rare gas mass spectrometry (RGMS)~\cite{rgms:Lindemann:2013kna}. Considering that natural krypton contains \isotope[85]{Kr} at the $2\times10^{-11}$ level~\cite{xenon1t:analysis:2019izt,XENON:2021fkt}, the RGMS measurements were used to constrain the associated background. The mean concentrations in SR1 and SR2 were $^{\text{nat}}\text{Kr}/\text{Xe}= (0.7\pm0.1)\;\text{ppt}$ and $^{\text{nat}}\text{Kr}/\text{Xe}= (1.2\pm0.2)\;\text{ppt}$ in mass, respectively. The latter arises from $^{\text{nat}}\text{Kr}/\text{Xe}= (0.7\pm0.2)\;\text{ppt}$ before the pump installation and $^{\text{nat}}\text{Kr}/\text{Xe}= (2.0\pm0.4)\;\text{ppt}$ thereafter. 

Several intrinsic backgrounds exhibited an explicit time-dependence. Neutron activation during calibrations led to backgrounds from the aforementioned \isotope[125]{I} and \isotope[125]{Xe} peaks, the metastable \isotope[131\text{m}]{Xe} peak at 163.9~keV with a half-life of 11.84~d, and the merged $\upgamma+\upbeta$ spectrum of \isotope[133]{Xe} with a half-life of 5.25~d. Due to the proximity to neutron calibrations, the activation level in SR1$_\text{b}$ was increased compared to SR1$_\text{a}$. The more frequent neutron calibrations in SR2 also led to a larger activation background contribution. Apart from \isotope[125]{I}, no constraints were placed on the rates of these backgrounds. 

A \isotope[83\text{m}]{Kr} peak at 41.5~keV was present in all datasets due to trace amounts of the parent isotope \isotope[83]{Rb} in the xenon recirculation system; its activity decreased with the \isotope[83]{Rb} half-life of $86.2\;\text{d}$. The decay of \isotope[83\text{m}]{Kr} is a two step isomeric transition~\cite{toi:firestone_shirley_1999} with energy depositions of $31.2\;\text{keV}$ and $9.4\;\text{keV}$ from conversion electrons. The $157\;\text{ns}$ half-life of the $9.4\;\text{keV}$ state leads to the merging of the corresponding S1s and S2s in XENON1T which produces the 41.5~keV peak. If both energy depositions can be distinguished, the events are removed by cuts. In a fraction of \isotope[83\text{m}]{Kr} events, the S1 from the 9.4\;keV transition was wrongly classified as an S2 which was not included in the energy reconstruction leading to a secondary lower-energy \isotope[83\text{m}]{Kr} peak. As outlined in Sec.~\ref{sec:edec:cuts}, a dedicated cut was developed in order to address this reconstruction artifact.

The elastic scattering of solar neutrinos off atomic electrons constituted a subdominant background compared to the intrinsic and material background components and was implemented as in~\cite{mc:xenon1t:XENON:2015gkh}. The backgrounds from detector materials and \isotope[136]{Xe} were implemented as outlined in Sec.~\ref{mc:common}.

\subsection{Fit method and parameters}\label{sec:edec:fit:method:pars}
Following the methodology of Sec.~\ref{fits:common}, a binned likelihood was constructed for each of the four measured energy spectra in the SR1$_\text{a}$ inner and outer FVs as well as the SR1$_\text{b}$ and SR2 cylinders, and used for a simultaneous fit of the signal and background models to the measured data. The SR1$_\text{a}^\text{in}$ and SR1$_\text{a}^\text{out}$ datasets were fitted with a 1 keV binning while the SR1$_\text{b}^\text{in}$ and SR2 data were fitted in 2 keV bins due to the lower exposure. Different binnings of 0.5 keV, 1.0 keV, 1.5 keV and 2.0 keV were tested for all datasets, but did not significantly affect the results. 
Relative differences in the obtained $2\upnu\text{ECEC}$ half-lives were less than $3\,\%$ and small compared to the systematic uncertainties.
The definition of the total likelihood for the signal and background model included 51 fit parameters. Of these, 29 parameters were constrained. Tables with constraints and best-fit values for all parameters can be found in the appendix. Twelve parameters were shared among all datasets. They included the $\dec$ signal and the background sources that were constant in time, such as detector construction materials, solar neutrinos and \isotope[136]{Xe}. The properties of the residual \isotope[83\text{m}]{Kr} reconstruction artifact, its position ($\mu_{^{83\text{m}}\text{Kr,misID}}$), width ($\sigma_{^{83\text{m}}\text{Kr,misID}}$) and relative frequency with respect to the main 41.5 keV peak ($f_{^{83\text{m}}\text{Kr,misID}}$), were also shared. Its constraints were derived from \isotope[83\text{m}]{Kr} calibration data.

The remaining 39 parameters were not shared among all datasets and can be grouped into four different categories. The first category includes homogeneously distributed backgrounds from \isotope[85]{Kr} and \isotope[214]{Pb}  which could be averaged in time over SR1 and SR2. 

The second category contains backgrounds that could not be averaged over the entire science run, but were sufficiently long-lived to distribute uniformly within the detector. The decay rates of the neutron-activated peaks of \isotope[125]{I} and \isotope[131\text{m}]{Xe}, as well as those for \isotope[125]{Xe} and \isotope[133]{Xe}, were dependent on their temporal proximity to the neutron calibrations. Due to its 16.9~h half-life, \isotope[125]{Xe} was only present in SR1$_\text{b}$. 

The third category contains the parameters describing the acceptance. These were constant over time in a given science runs, but differ in the inner and outer volumes of the detector. The SR1 acceptances for the inner and outer volumes were parameterized by linear functions of reconstructed energy that were fitted to \isotope[220]{Rn} calibration data in order to derive the parameter constraints. A constant parameterization was used for SR2. 

The fourth and last category consists of parameters that were both time- and position-dependent, so they were fitted individually for each dataset. For the \isotope[83\text{m}]{Kr} decay rate $A_{^{83\text{m}}\text{Kr}}$, the time dependence originated from the decay of a \isotope[83]{Rb} contamination with a half-life of 86.2~d. The spatial dependence is a feature of the event reconstruction. Since \isotope[83\text{m}]{Kr} undergoes a two-step decay, events only appeared as a 41.5~keV peak if the S1s and S2s from the subsequent isomeric transitions were merged by the data processor. 
While the S2s were always identified as a single signal in the bulk of the detector due to their $\mathcal{O}(\upmu\text{s})$ width, the S1s could sometimes be distinguished depending on their separation in time and the decay position in the TPC. For the same \isotope[83\text{m}]{Kr} activity in the inner and outer volumes of the TPC, this led to different 41.5 keV peak areas. 

The remaining parameters implemented systematic uncertainties on the energy resolution and reconstruction in the fit. The energy resolutions $\sigma_E$ of the monoenergetic peaks with true energy $\mu_E$ were parameterized as~\cite{Aprile:2020yad}
\begin{align}
  \sigma_E = a_\text{res}\times\sqrt{\mu_E} + b_\text{res}\times \mu_E, 
\end{align}
with the constrained fit parameters $a_\text{res}$ and $b_\text{res}$ for each dataset. The SR1 parameters were constrained using the parametrization from~\cite{Aprile:2020yad}. For the SR2 resolution, fits of monoenergetic calibration lines were used. The simulated spectra were smeared prior to the fit with the same function. 

In order to account for a possible bias in the energy reconstruction, the fitted energy $E_\text{fit}$ for each signal and background component could be shifted from the simulated energy $E$ by adding a linear energy-dependent shift in the fit
\begin{align}
  E_\text{fit}(E) = E+ a_\text{shift}\times(E-b_\text{shift}).
\end{align}
Here, $a_{\text{shift}}$ and $b_{\text{shift}}$ refer to fit parameters modeling this energy shift. 
The shifts were applied independently for all datasets in order to account for possible different behaviors in the FVs and for possible temporal drifts. This is obvious for SR2 where the energy reconstruction and resolution parameters were different from those determined for SR1. For SR1 these parameters were averaged over the entire science run and only determined in the inner detector volume. Since no independent calibration data is available over the whole duration of SR1 that could be used to formulate constraints, $a_\text{shift}$ and $b_\text{shift}$ were left unconstrained for all datasets. 

\subsection{Fit results}\label{sec:2vecec_results}
    \begin{figure*}[t]
      \centering
      \includegraphics[width=\textwidth]{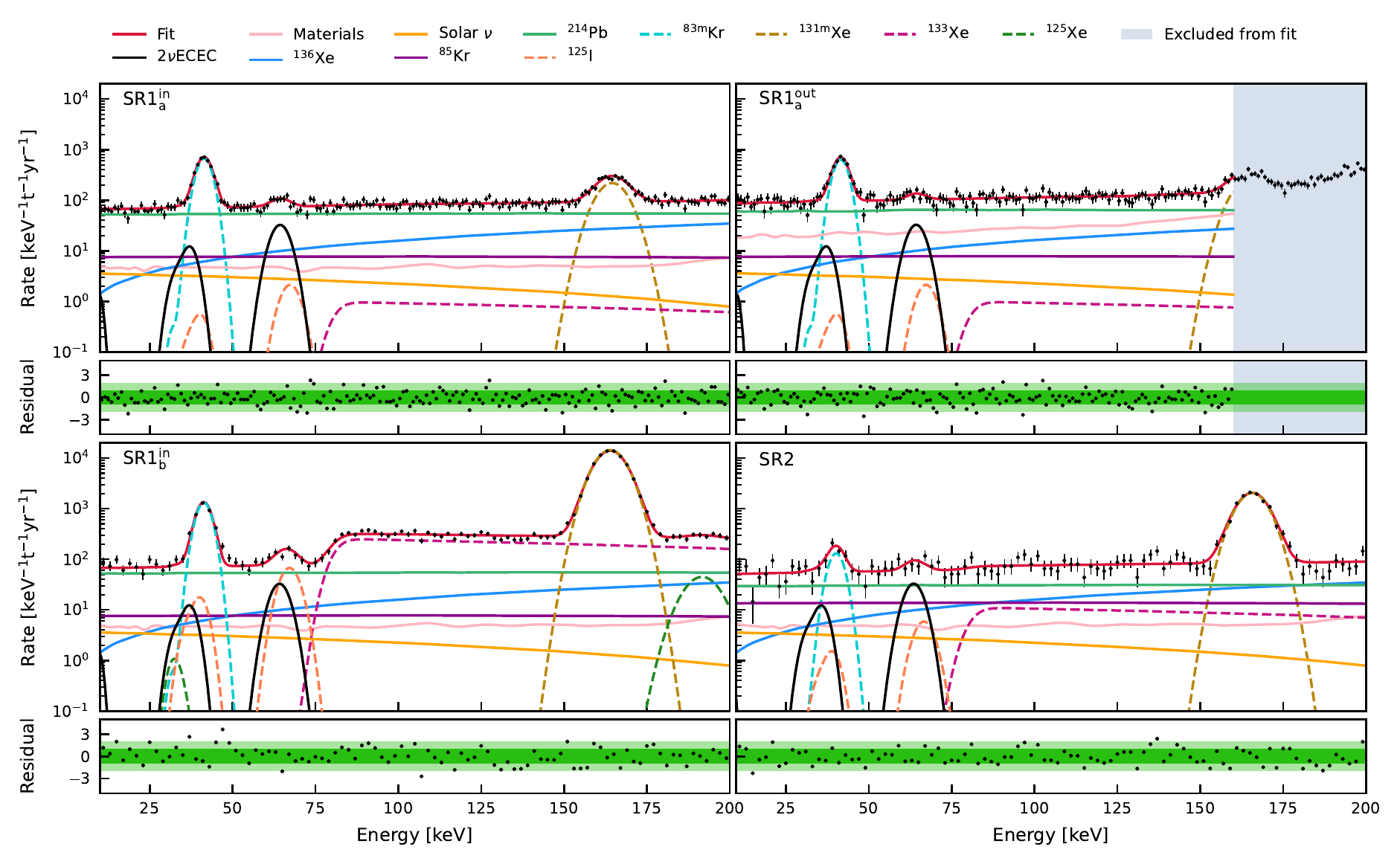}
      \caption{Fit of the combined signal and background model to the measured data with $\chi_\lambda^2/\text{ndf} = 517/508= 1.02$. Uncertainties on the data points are $68.3\,\%$ Feldman \& Cousins confidence intervals~\cite{feldman:cousins:Feldman_1998} on the number of counts per bin scaled with the exposure. The four panels show SR1$_\text{a}^\text{in}$ (top left), SR1$_\text{a}^\text{out}$ (top right) and SR1$_\text{b}^\text{in}$ (bottom left) as well as SR2 (bottom right).
      The sum spectrum is indicated by the solid red line.
      Background sources constant in time are shown as solid lines while those that vary over time are shown as dashed lines. The $2\upnu$ECEC signal peaks are indicated by solid black lines. The residuals were calculated from the square-roots of the $\chi^2_\lambda$ summands. 
      }
      \label{fig:edec:fit_sr1_sr2}
    \end{figure*}
The best-fit combined signal and background model is shown in Fig.~\ref{fig:edec:fit_sr1_sr2}. The reduced $\chi^2_\lambda/n_\text{dof}$ of the fit is $517/508 = 1.02$. This considers 530 data points, 51 fit parameters and 29 pulls from constrained parameters with the sum of squared pulls being $\Sigma=7.6$. The smallness of the pull contributions is attributed to the fact that a pull either does not originate from a statistical confidence interval or that the respective parameter is more strongly constrained by the auxiliarly constraint than by the science data. This is the case for the parameters describing the remaining part of the \isotope[83\text{m}]{Kr} reconstruction artifact, as the total area of the peak is more than two orders of magnitude below the signal rate. Therefore, the smallness of $\Sigma$ does not indicate a problem with the fit. The reduced $\chi^{2\prime}_\lambda/n_\text{dof}^\prime$ just from residuals and excluding pulls is $509/479=1.06$.

The spectra in SR1$_\text{a}^\text{in}$ and SR1$_\text{a}^\text{out}$ are featureless except for the monoenergetic peaks from the $2\upnu\text{ECEC}$ signal, \isotope[83\text{m}]{Kr} and neutron-activation. The spectrum in the outer volume has a larger slope due to the increased material background contribution. As expected, SR1$_\text{b}^\text{in}$ and SR2 exhibit larger neutron-activated peaks as well as the step from the merged $\upbeta+\upgamma$ signature of \isotope[133]{Xe}. The contribution from neutron activation in SR2 is lower than in SR1$_\text{b}$, due to the selection of datasets at least 50~d away from neutron calibrations, but higher than in SR1$_\text{a}$. The smallest \isotope[83\text{m}]{Kr} peak is found in SR2 due to the largely decayed \isotope[83]{Rb} contamination. 

The best-fit activity concentration of \isotope[214]{Pb} in SR1 is close to the XENON1T target value of ${~10\;\upmu\text{Bq}/\text{kg}}$ with ${^{214}\text{Pb}_\text{SR1}=(9.3\pm0.4)\;\upmu\text{Bq}/\text{kg}}$.
For SR2 it is $^{214}\text{Pb}_\text{SR2}=(5.3\pm0.8)\;\upmu\text{Bq}/\text{kg}$. 

\begin{figure}
  \centering
  \includegraphics[width=\linewidth]{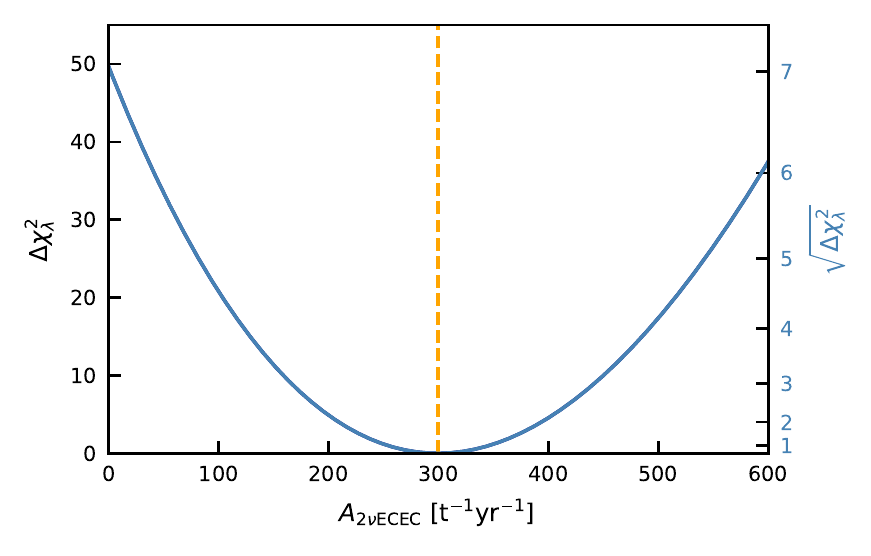}
  \caption{The $\chi^2_\lambda$ profile of the double-electron capture decay rate $A_{2\upnu\text{ECEC}}$. The minimum is indicated by the dashed orange line. The left $y$-axis gives the $\Delta\chi^2_\lambda$ between the best-fit rate at the minimum and the scanned rate. The right $y$-axis marks the significance level for excluding a null-result. The significance according to the profile is $7.0\,\upsigma$.}
  \label{fig:edec:profile_sr1_sr2}
\end{figure}
The significance of the $2\upnu\text{ECEC}$ signal was derived from the $\chi^2_\lambda$ profile of $A_{2\upnu\text{ECEC}}$ as shown in Fig.~\ref{fig:edec:profile_sr1_sr2}. The best-fit double-electron capture rate is 
\begin{equation}
   A_{2\upnu\text{ECEC}}=(300\pm50)\;\text{t}^{-1}\text{yr}^{-1} .
\end{equation}
The difference ${\Delta\chi^2_\lambda=49.4}$ between the best-fit rate and a null result yields a significance of $7.0\,\upsigma$ for the presence of a double-electron capture signal. This marks the first significant detection ($>5\,\upsigma$) of a two-neutrino double-electron capture in any isotope. Moreover, it is the first measurement of this process that leverages the signatures of higher-shell KL-, KM-, KN- and LL-captures.

The resulting $2\upnu\text{ECEC}$ half-life using Eq.~(\ref{eq:0vbb:halflife:calc}) is
\begin{align}
  T_{1/2}^{2\upnu\text{ECEC}}= (1.1\pm0.2_\text{stat}\pm0.1_\text{sys})\times10^{22}\;\text{yr}.
\end{align}
The systematic uncertainty has four individual contributions given in Tab.~\ref{tab:edec:systematics}: the cut acceptance, the exposure, the \isotope[124]{Xe} isotopic abundance, and the theoretical uncertainties on the relative fraction of KK-, KL and higher-shell captures. The first three contributions were calculated with Gaussian uncertainty propagation in Eq.~(\ref{eq:0vbb:halflife:calc}) while the fourth was derived by comparing fit results using two different signal models. The uncertainty on the acceptance was obtained by calculating the average acceptance over the entire energy range for each dataset. The difference of the exposure-weighted sum of these average values from unity was taken as the systematic uncertainty. The uncertainty on the exposure was obtained from the FV uncertainties of each dataset that were multiplied with the corresponding live times and added in quadrature. The total uncertainty on the \isotope[124]{Xe} isotopic abundance was obtained by adding the statistical and systematic uncertainties in quadrature. 

As stated in Tab.~\ref{tab:edec:fractions}, approximate values for the double-electron capture fractions from different shells were used. These were calculated from the overlap of the nuclear and electronic wave functions. In order to determine the systematic uncertainty arising from this approximation, the full analysis was also carried out with a simplified model including only captures from the K- and L1-shells. As the scaling of the $\dec$ model is predominantly determined by the KK-peak, the increased capture fraction in the simplified model lead to a half-life that is $6.3\,\%$ longer. The absolute difference is used as the systematic uncertainty.
\begin{table}[t]  
\centering
  \begin{center}
      \caption{Systematic uncertainties on the $\dec$ half-life. The total systematic uncertainty was obtained by adding the individual components in quadrature and rounding to the first digit.}
      \label{tab:edec:systematics}
      {
      \tabcolsep=1.00mm
      \begin{ruledtabular}
      \begin{tabular}{l c c}
      Contribution & Uncertainty\;[$10^{22}\;\text{yr}$] & Relative\;[$\%$]\\
      \colrule
    Acceptance & $0.05$ & $4.5$ \\
    Exposure & $0.003$ & $0.3$\\
    \isotope[124]{Xe} abundance & $0.02$ & $1.8$\\
    Capture fractions & $0.07$ & $6.3$\\
    \colrule
    Total & $0.1$ & $9.1$\\
      \end{tabular}
      \end{ruledtabular}}
  \end{center}
\end{table}

\subsection{Comparison with theory and other experiments}\label{sec:2vecec_comparison_results}
\begin{figure}
  \centering
  \includegraphics[width=\linewidth]{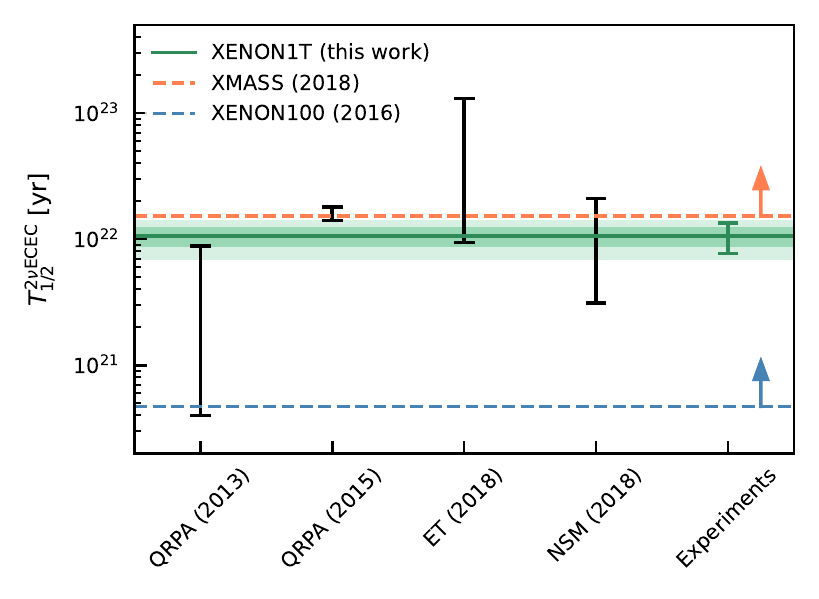}
  \caption{Comparison of the $2\upnu$ECEC half-life with theoretical predictions and the experimental $90\,\%\;\text{CL}$ lower limits from XMASS~\cite{xmass:Abe:2018gyq} (dashed orange) and XENON100~\cite{XENON:2016jyx} (dashed blue). As previous results considered a signal model with the double-K transition only, the lower limits were scaled down with the double-K capture fraction from this work. The updated central value of the measured half-life is shown as the solid green line. The $1\,\sigma$ and $2\,\sigma$ statistical uncertainty bands are indicated in green. The green uncertainty bar indicates $1\,\sigma$ of the sum of the statistical and the total systematic uncertainty. Four half-life ranges from nuclear structure calculations~\cite{Suhonen:2013rca, nme-PhysRevC.91.054309, CoelloPerez:2018ghg} are indicated in black. The NSM and ET predictions were scaled with the double-K capture fraction while the QRPA ones are already given for the total $\dec$ half-life. The outer bounds of the half-life ranges predicted by all models are within twice the statistical uncertainty of our result.}
  \label{fig:edec:halflife}
\end{figure}
The new result can be compared with the previously measured $2\upnu$KK-half-life ${T}_{1/2}^{2\upnu\text{KK}}= (1.8\pm0.5_\text{stat}\pm0.1_\text{sys})\times10^{22}\;\text{yr}$ from~\cite{XENON:2019dti}. We use the KK-capture fraction of $72.4\,\%$ to compute $T_{1/2}^{2\upnu\text{KK}}= (1.5\pm0.3_\text{stat}\pm0.1_\text{sys})\times10^{22}\;\text{yr}$ for this work. The datasets partially overlap, so the statistical uncertainties are correlated. However, the analyses used different data processor versions, cuts and energy reconstructions. Consistency checks of both results were carried out using the 0.68~$\text{t}\times\text{yr}$ data contained in both analyses.
It was found that the small difference between both results for $T_{1/2}^{2\upnu\text{KK}}$ can be accounted for by the updated signal model, the improved energy reconstruction and the larger cut acceptance in this work, together with the independent systematic uncertainties as well as the $33\,\%$ larger exposure. 

Fig.~\ref{fig:edec:halflife} compares the measured half-life with the most recent calculations from four theoretical approaches. Due to the shorter half-life compared to the former XENON1T analysis~\cite{XENON:2019dti}, the agreement with the QRPA (2013) calculation~\cite{Suhonen:2013rca} is improved. The value range from QRPA (2015)~\cite{nme-PhysRevC.91.054309} is consistent with our new result at the $2\,\sigma$ level. Both the ET and the NSM calculations are compatible with our new result~\cite{CoelloPerez:2018ghg}.
While the central value of the first XENON1T result was less than $1\,\sigma$ below the $90\,\%\;\text{CL}$ lower limit of XMASS, the new result is approximately $2\,\sigma$ below the XMASS limit.

Future xenon-based detectors with lower backgrounds and larger exposures will further probe $\dec$ to improve experimental constraints on NME calculations for proton-rich nuclides. The best-fit rate from this work would result in XENONnT detecting approximately 6000 double-electron capture events in its projected 20~$\text{t}\times\text{yr}$ total exposure. With a reduction in background by a factor of $\sim6$~\cite{xenonnt_mc:Aprile_2020}, the half-life could be measured with a precision at the few-percent level and the relative capture fractions could be investigated. In this regard theoretical input on the relative capture fractions as well as the double-hole energies is needed. Moreover, with more exposure and less background the $\dec{}$ can be used as an ideal internal energy calibration source, and the remaining two-neutrino and hypothetical neutrinoless decays of \isotope[124]{Xe}~\cite{Xe124:WITTWEG2020} could become accessible.

\section{Search for $^{136}$Xe neutrinoless double-$\boldsymbol{\upbeta}$ decay in XENON1T}
\label{sec:1t_0vbb}
In contrast to dedicated \doublebeta{} experiments with xenon inventories enriched in \isotope[136]{Xe}~\cite{exo200:PhysRevLett.123.161802, KamLAND-Zen:2022tow}, the isotopic composition of the XENON1T target was close to that of natural xenon with an abundance of \isotope[136]{Xe} as mentioned in Sec.~\ref{mc:common}.
With a tonne-scale fiducial mass and two years of measurement time, an isotope exposure of $36.16\;\text{kg}\times\text{yr}$ was achieved, approaching exposures of dedicated experiments with enriched targets. 
The data used in this analysis are a subset of the SR1 dataset introduced in Sec.~\ref{sec:common_info}.
Data periods when the neutron generator was in the water tank close to the cryostat were removed from the data selection to avoid an elevated high-energy $\upgamma$-ray background level from thorium and radium decay chain isotopes in the neutron generator's materials.
The total live time of the dataset is 202.7~days.
The analysis was performed in the energy range between 1600~keV and 3200~keV in order to include multiple $\upgamma$-peaks that helped to constrain material background components and covered the endpoint region of the \isotope[214]{Bi} $\upbeta$-spectrum.
A blinding cut between 2300~keV and 2600~keV was applied to the dataset.

\subsection{Event selection}
\label{1t:ev_selection}
The event selection criteria for signal-like interactions were developed on the blinded science data as calibration sources with energies close to $Q_{\upbeta\upbeta}$ were not available in XENON1T.
The applied cuts are based on those from~\cite{Aprile:2019bbb} and were adapted to higher energies.
Firstly, data quality criteria were applied to remove events in coincidence with muon veto triggers and data acquisition busy periods.
Periods with light emission in the PMTs~\cite{Barrow_2017}, causing abnormal data rates, were also removed.
The $\doublebeta{}$ signal is expected to be an SS interaction, while events involving Compton scattering will typically be MS interactions.
Thus, SS events were selected by rejecting events with a second S2 whose size, width, and PMT hit-pattern were compatible with the S1, such that the secondary S2 and the primary S1 would form a valid event.
A multi-S1 cut, based on the size of the second largest S1 in an event, rejected interactions with multiple S1s originating from pile-up.
One source of pile-up is \isotope[214]{BiPo} decay, discussed in Sec.~\ref{mc:common}, with two subsequent decays occurring in the same event.

Two different position reconstruction algorithms, a neural network using TensorFlow~\cite{tensorflow2015-whitepaper} and an algorithm using a fit of the S2 hit pattern on the top PMT array~\cite{Aprile:2019bbb}, were required to give compatible results.
This removed events close to the edge of the TPC or in regions where non-functioning PMTs were located. We also required the reconstructed position to be compatible with the observed S2 hit-pattern of the top PMT array. Cuts based on the fraction of light detected by the top and bottom PMT arrays, for both the S1 and S2 signals, were effective at removing events from energy depositions in the gaseous xenon layer.
ERs were identified by the ratio of the S1 to the S2 signal size with 98\,\% efficiency. Finally, we applied a cut requiring the S2 width to be compatible with the expected diffusion of drifted electrons from the reconstructed depth $(z)$ in the TPC. 

The individual cut acceptances were determined with three different techniques.
The exposure loss from data quality criteria was factored into the live time.
Cuts whose acceptance was tested on controlled samples of data have fixed acceptances corresponding to the fraction of the parameter distribution passing the cut.
The rest of the cut acceptances were determined iteratively by comparing the number of remaining events after a set of cuts with the number of events after applying the same cut set, except for the one under investigation.
The combined cut acceptance was then determined by multiplication of the individual cut acceptances per energy bin.
The result was interpolated with a quadratic spline weighted by acceptance uncertainties at each data point~\cite{2020SciPy}.
This provided a continuous acceptance parametrization over the full energy range of interest. Since it was not possible to differentiate between removed signal and background events, the iterative method provided only a lower limit on the signal acceptance. In the blinded region this was approximately flat and extrapolated to be $>88\,\%$ at $Q_{\upbeta\upbeta}$.
An upper limit of 97.5\,\% was determined by considering only cuts with a fixed acceptance as outlined above. 
In the later fit of the signal and background models, discussed in Sec.~\ref{subsubsec:fit_blind}, the acceptance was allowed to float between the lower and upper limits.

\begin{figure}[t]
	\centering
	\includegraphics[width=\linewidth]{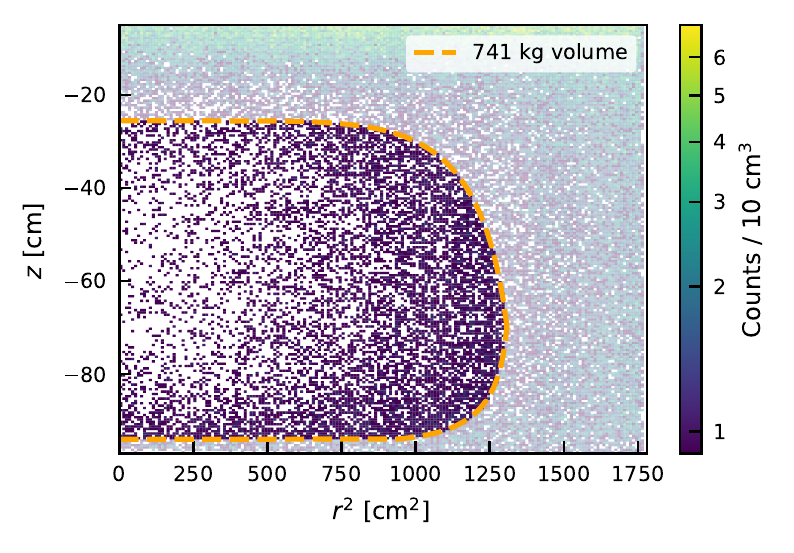}
 \caption{Reconstructed position distribution of events in $4\,\sigma_E$ regions around the \isotope[214]{Bi} and \isotope[208]{Tl} peaks at 2204.1\;keV and 2614.5\;keV, respectively. The dashed orange line shows the optimized 741\;kg FV.
 }
	\label{fig:fv_binning}
\end{figure}

An inner FV was selected based on a sensitivity figure of merit $S_\text{vol}$ in order to maximize the signal to noise ratio:
\begin{align}
 S_\text{vol} \propto \frac{m}{\sqrt{B}},
 \label{eq:0vbb:fom}
\end{align}
where $m$ is the target mass which scales linearly with the number of signal events and $B$ is the number of expected background counts.
Two control regions close to $Q_{\upbeta\upbeta}$ in the science data were used for this study.
They were defined as $\pm 4\,\sigma_E$ intervals around the \isotope[214]{Bi} and $^{208}$Tl peaks at 2204.1\;keV~\cite{a214:WU2009681} and 2614.5\;keV~\cite{a208:MARTIN20071583}, respectively, excluding the data in the blinded region. The resolution at these energies was $\sigma_E/E=0.8$\%.
The TPC's active volume was binned in a $9\times9$ grid in squared radius $r^2$ and depth $z$ containing equal masses. 
The sensitivity figure of merit $S_\text{vol}$ was computed in each bin using the sum of events from both control regions.
The resulting grid of sensitivity values was then smoothed to 100 contour levels of the same $S_\text{vol}$, which was fitted with two semi-superellipsoid functions.
The maximum allowed depth of the FV was $-94$\;cm in order to avoid TPC regions with possible field distortion close to the cathode region.
Finally, $S_\text{vol}$ was computed for the volume enclosed in the fitted contours. This resulted in an optimal FV containing $(741\pm9)\;\text{kg}$, shown in Fig.~\ref{fig:fv_binning} with the event distributions in the control regions.
The volume is shifted towards the bottom of the detector due to the presence of more material at the top of the TPC and less shielding from the xenon in gaseous phase. 

\subsection{Background model}
\label{sec:bkg_1t}
The background model for the $\doublebeta{}$ search accounts for backgrounds from intrinsic and external sources. The background model was validated by a fit to the science dataset between 1600\;keV and 3200\;keV, excluding the blinded region.

The dominant background in the ROI, defined as the $2\,\sigma$ region around $Q_{\upbeta\upbeta}$, was due to $\upgamma$-rays from trace amounts of radioactive isotopes in detector components, as already discussed in Sec.~\ref{mc:common}. The main contributors were the late parts of the primordial \isotope[238]{U} and \isotope[232]{Th} decay chains as well as \isotope[60]{Co}.
Most notable were the full absorption peak of the $2447.9$~keV $\upgamma$-ray of \isotope[214]{Bi}~\cite{a214:WU2009681} and Compton scatters from the $2614.5$~keV \isotope[208]{Tl} line~\cite{a208:MARTIN20071583}.
Additionally, a peak at $2505.7$~keV originating from two \isotope[60]{Co} $\upgamma$-rays at $1173.2$~keV and $1332.5$~keV~\cite{a60:BROWNE20131849}, detected in coincidence as an SS event, was expected.
The early parts of the primordial decay chains did not contribute to the background in the ROI, but appeared in the 1600--3200~keV fit range.

The $\twonubeta$ decay of \isotope[136]{Xe} and the $\upbeta$-decay of \isotope[214]{Bi}, entering the FV by radon emanation, were considered as intrinsic background sources. Details regarding their contributions are discussed in Sec.~\ref{mc:common}. Since the \isotope[214]{BiPo} tagging efficiency was not known from external measurements, the intrinsic \isotope[214]{Bi} background component was not constrained in the fit.
The measured spectrum is continuous up to the endpoint at $(3270\pm11)$\;keV.
Spectral features occur where the emitted $\upbeta$ and subsequent $\upgamma$-rays are merged into a single energy deposition.
Decays occurring in the LXe shell outside of the TPC could not be tagged since the $\upalpha$ from the \isotope[214]{Po} decay did not enter the active region of the detector.
Only the $\upgamma$-rays following the $\upbeta$-decay could be registered inside the active volume. Thus, they had to be treated separately from the TPC contribution.
The background fit constraint of $(10\pm5)\;\upmu\text{Bq}/\text{kg}$ for the \isotope[214]{Bi} activity concentration was informed by $\upalpha$-decay measurements~\cite{xenon1t:radon:2020fbs}.

Neutron-activated $^{137}$Xe and ERs induced by $^{8}$B solar neutrinos are negligible compared to the material background in XENON1T and were not considered here.

\subsection{Fit to the blinded data and sensitivity}
\label{subsubsec:fit_blind}
The methodology for fitting and limit setting was introduced in Sec.~\ref{fits:common}.
The set of nuisance parameters comprises scaling factors for all simulated backgrounds.
Additionally, we considered a combined cut acceptance parameter $\epsilon$, which was allowed to move between the lower bound of 88\% and the upper bound of 97.5\% in the ROI within $1\,\sigma$ of its Gaussian constraint, as mentioned in Sec.~\ref{1t:ev_selection}.
\begin{figure}[t]
	\centering
	\includegraphics[width=\linewidth]{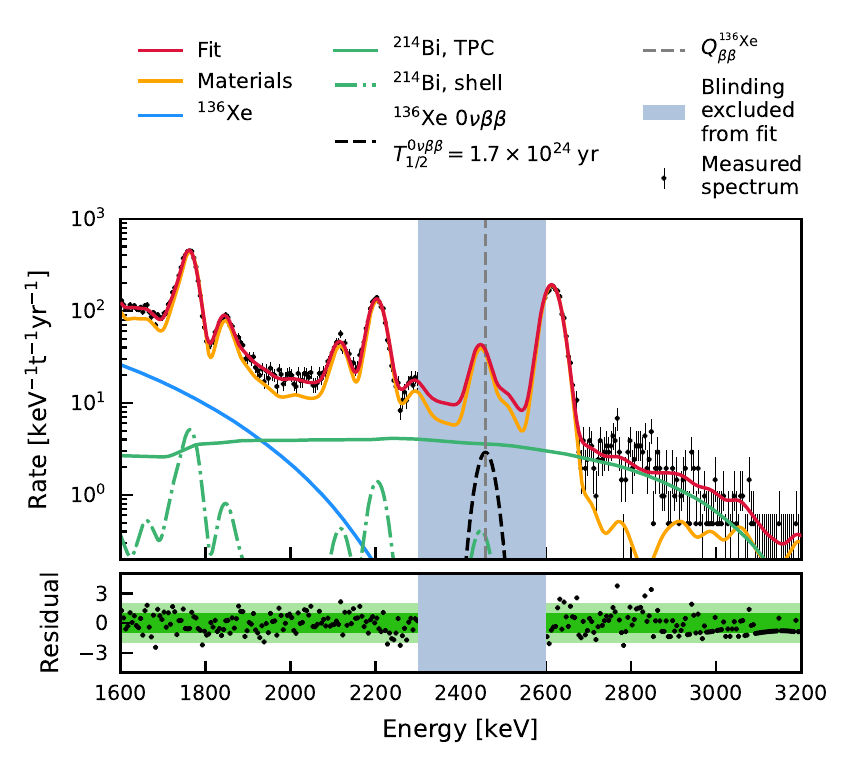} 
	\caption{Pre-unblinding data (black) and background model fit (red) between 1600\;keV and 3200\;keV. The background from materials (gold), the intrinsic \isotope[136]{Xe} (blue), and the \isotope[214]{Bi} inside the active volume of the TPC (green) and in the LXe shell (dashed green) are also displayed. The bottom panel shows the residuals with the 1$\sigma$ and 2$\sigma$ bands. Due to low statistics in the measured data above 2800~keV the residuals over the entire energy range were normalized with the square-root of the expected counts from the best-fit model. A hypothetical $0\upnu\upbeta\upbeta$ peak at the exclusion sensitivity of $1.7\times10^{24}\;\text{yr}$ is shown in the blinded region (black dashed line).}
	\label{fig:xenon1t:0vbb:blinded}
\end{figure}
\begin{figure}[t]
	\centering
	\includegraphics[width=\linewidth]{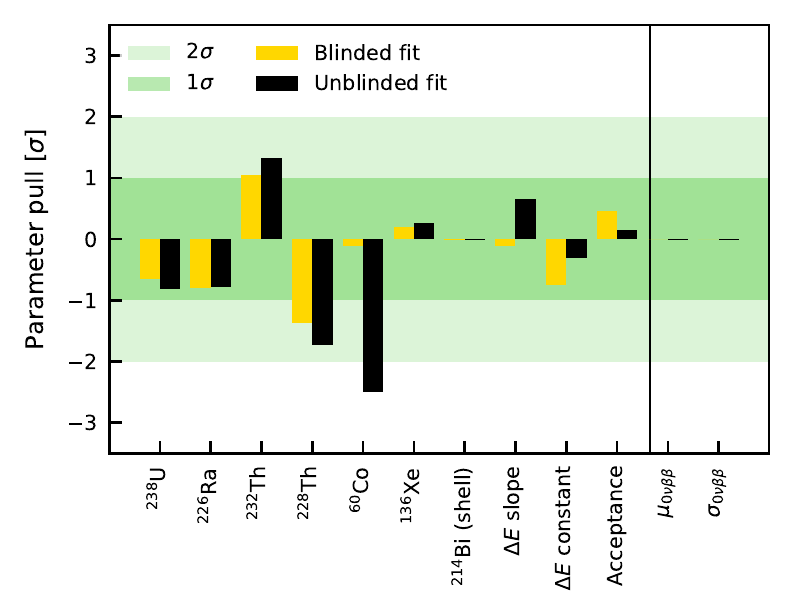} 
	\caption{Parameter pulls of the fit for the blinded (yellow) and unblinded data (black) in units of the constraint uncertainties $\sigma$. The parameters describing the $\doublebeta$ peak, $\mu_{0\upnu\upbeta\upbeta}$ and $\sigma_{0\upnu\upbeta\upbeta}$, are only present in the fit to the unblinded data.}
	\label{fig:xenon1t:0vbb:pulls}
\end{figure}
The positions of high-energy peaks agree with the expected energy within $\pm0.5\,\%$~\cite{Aprile:2020yad}.
In order to correct for the remaining residual energy shift due to systematic uncertainties, we included two fit parameters, $\Delta E\, _{\text{slope}}$ and $\Delta E\, _{\text{offset}}$. The energies of the simulated events $E_\text{MC}$ were then allowed to move as
\begin{align}
 E_\text{fit} = E_\text{MC} + \Delta E,
\end{align}
where $\Delta E$ was parametrized as
\begin{align}
 \Delta E = \underbrace{(1.5\pm0.2)\times 10^{-3}}_{\Delta E\, \text{slope}} \times E_\text{MC} \underbrace{- (4.4\pm0.3)}_{\Delta E\, \text{offset}}\;\text{keV},
\end{align}
with constrained parameters for slope and offset.

Fig.~\ref{fig:xenon1t:0vbb:blinded} shows the fit to the blinded data, which is well described with $\chi^2_\lambda/\text{ndf} = 311/259=1.20$.
In the high-statistics region below 2800~keV, the residuals are symmetric and centered around zero with a standard deviation of $\sigma_\text{res} = 1.05$.
In the low-statistics region above 2800~keV, the fit lies mostly above the measured data leading to negative residuals and an asymmetric distribution, as the fit function can only predict rates larger than or equal to zero.
The parameter pulls are shown in Fig.~\ref{fig:xenon1t:0vbb:pulls}.
None of the pulls for the blinded fit exceeds $2\,\sigma$ and the sum of the squared pulls is $4.6$.
The pull on \isotope[60]{Co} is close to zero since its double-$\upgamma$ peak is located in the blinded region.
Due to the degeneracy with the \isotope[226]{Ra} spectrum and its small background contribution, the \isotope[214]{Bi} component in the LXe shell outside of the TPC is less sensitive to the data than to its constraint and is not pulled away from the expected value.
The parameters for the uranium and thorium chains are within the expected range.
No notable pulls on the systematic uncertainty parameters are observed. The acceptance parameter prefers a value close to the lower bound. 

In order to compute the sensitivity, a $\doublebeta$ signal was added to the background model as a Gaussian peak.
Its mean ${\mu_{0\upnu\upbeta\upbeta}=(2457.8\pm0.4)\;\text{keV}}$ is given by the Q-value~\cite{PhysRevLett.98.053003,q-value:McCowan:2010zz}, the standard deviation $\sigma_{0\upnu\upbeta\upbeta}=(19.7\pm0.3)\;\text{keV}$ is given by the energy resolution. The SS fraction $\epsilon_\text{SS}=90.3\,\%$ of signal events was determined with MC simulations. Initial momenta for $10^6$ electron pairs were generated with DECAY0~\cite{dec-tretyak}, their tracks were propagated with Geant4 and clustered based on the $z$-separation of subsequent energy depositions.

The expected sensitivity for setting a lower limit on the \doublebeta{} half-life $T_{1/2}^{\doublebeta{}}$ of \isotope[136]{Xe} was determined using toy-MC simulations as outlined in~Sec.~\ref{fits:common}. We derive a median upper limit on the decay rate with $A_{\doublebeta}<144\;\text{t}^{-1}\text{yr}^{-1}$ at $90\,\%\;\text{CL}$.
Using Eq.~(\ref{eq:0vbb:halflife:calc}), the expected sensitivity on the blinded data is
\begin{align}
 T_{1/2\text{,\;expected}}^{0\upnu\upbeta\upbeta}>1.7\times 10^{24}\;\text{yr}\; \text{at}\; 90\,\%\;\text{CL}.
 \label{eq:halflive_from_fit}
\end{align}

\subsection{Post-unblinding changes and final results}
\label{subsec:postunblinding}
After unblinding the events in the $\doublebeta$ ROI, an unexpected excess of events was observed around 2550~keV, well above the Q-value. This excess increased over time and was localized at the edges of the active volume. This indicated that an external background source progressively leaked into the selected data. Our investigation pointed to a class of MS events that were not rejected by the previously defined cuts, but that were misidentified as SS events. These events had a secondary S2 signal which was smaller than and temporally close to the main S2, and likely caused by multiple Compton scatters of a single $\upgamma$-ray. As the secondary S2 contained a part of the total deposited energy, the misidentified population was reconstructed at a lower energy with respect to the $\upgamma$-peak. The effect was present for all peaks in the ROI, but only the \isotope[208]{Tl} peak with its rising edge in the blinded region was large and isolated enough to significantly affect the $\doublebeta$ search. 
\begin{figure}
	\centering
	\includegraphics[width=\linewidth]{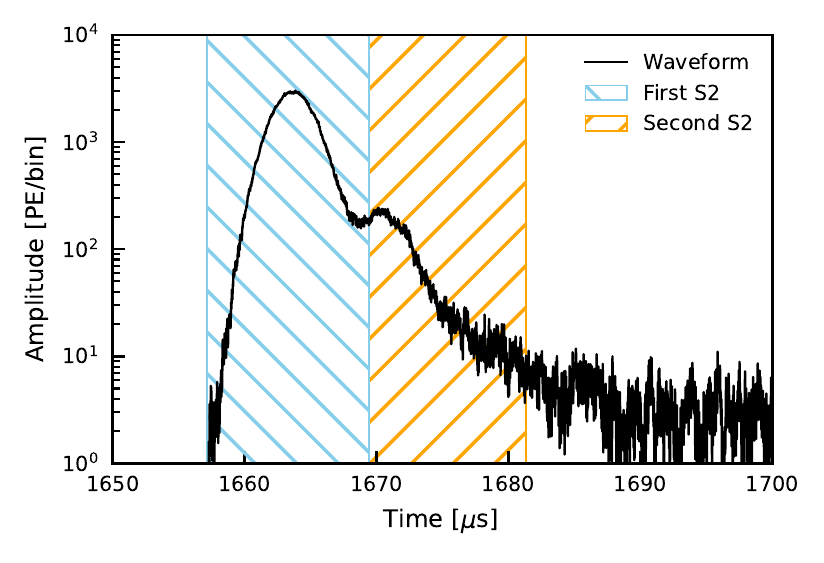} 
	\caption{Example of an MS event waveform only rejected by the post-unblinding cut. The primary S2 is indicated by the blue hatched region, while the secondary S2 due to Compton scattering is indicated in orange. The waveform of an SS event should exhibit a single S2 peak such as the one indicated in blue. Without the stricter cut this event was wrongly identified as SS. However, the SS energy was reconstructed only from the main S2 and the energy information of the smaller S2 after the main S2 was not considered. Accordingly, the event was reconstructed at a lower energy than deposited in the event.}
	\label{fig:wf_rejected}
\end{figure}

\begin{figure}
	\centering
	\includegraphics[width=\linewidth]{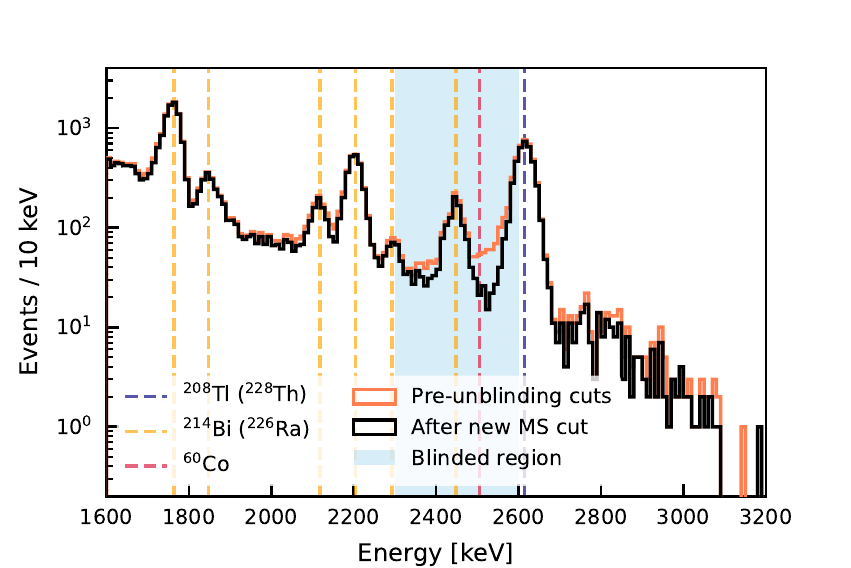} 
	\caption{Comparison of the full energy spectrum with the original cut set (orange) and after the addition of the stricter MS cut (black).}
	\label{fig:xenon1t:0vbb:post_unblinding_cut}
\end{figure}

The time-dependence of the effect is assumed to originate from increased PMT afterpulsing rates over time: MS events were identified based on the peak area and the top PMT hit pattern of the second largest S2 signal that was found in an event. PMT afterpulses that occurred in coincidence with the S2 altered the hit pattern as well as the signal size. Although the original MS classification accounted for the growth of afterpulsing with time, a stricter cut was needed to remove pathological waveforms such as the one shown in Fig.~\ref{fig:wf_rejected}. A post-unblinding cut was introduced, based on the peak area of the secondary S2 and its top PMT array hit pattern. The effect of the cut is shown in Fig.~\ref{fig:xenon1t:0vbb:post_unblinding_cut}.

The acceptance of the new cut was determined by comparing the number of events contained in the \isotope[208]{Tl} peak and multiple \isotope[214]{Bi} peaks outside of the previously blinded region before and after the new cut. MS events had a smaller reconstructed energy in the main S2 signal, as part of the energy was deposited in the subsequent S2 peaks. Thus, the centers of the $\upgamma$-lines -- especially of \isotope[208]{Tl} as the highest energy line -- provided pure samples of SS. We found an acceptance of $(97\pm2)\,\%$ and updated the total cut acceptance accordingly. 

\begin{figure}
	\centering
	\includegraphics[width=\linewidth]{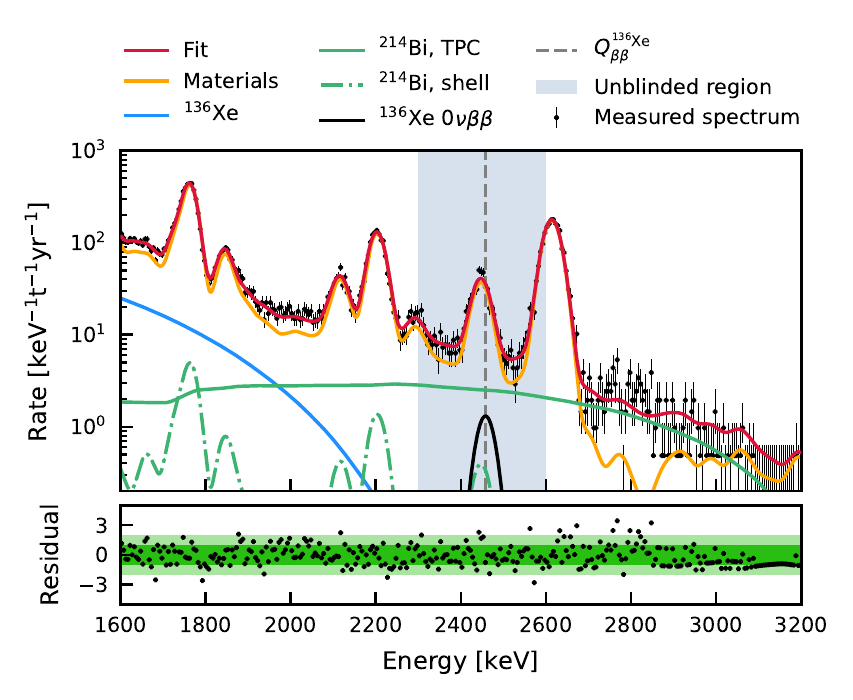}
	\caption{Final data (black) and background model fit (red) between 1600\;keV and 3200\;keV with post-unblinding changes.}
	\label{fig:xenon1t:0vbb:unblinded}
\end{figure}
\begin{table*} 
\centering
 \begin{center}
 \caption{Expected and best-fit background event counts in the $2\,\sigma_E$ ROI around $Q_{\upbeta\upbeta}$. The best-fit numbers are given for the fit to the data before unblinding, and for the fit to the unblinded data with the post-unblinding cut on multi-site events. The \isotope[60]{Co} upper limit is given at $90\,\%\;\text{CL}$. The expected \isotope[214]{Bi} events in the TPC and LXe shell are given for an assumed activity concentration of $(10\pm5)\;\upmu\text{Bq}/\text{kg}$.}
   \label{tab:0vbb:backgrounds}
   {
   \tabcolsep=1.00mm
   \begin{ruledtabular}
\begin{tabular}{l c c c}
                   Source &   Expected events & Blinded fit events & Unblinded fit events \\
\colrule
        \isotope[226]{Ra} &    $1200 \pm 600$ &        $751 \pm 9$ &         $760 \pm 10$ \\
        \isotope[228]{Th} &      $130 \pm 70$ &        $123 \pm 3$ &          $119 \pm 3$ \\
         \isotope[60]{Co} &       $70 \pm 30$ &        $70 \pm 20$ &             $ < 11 $ \\
        \isotope[136]{Xe} &     $4.5 \pm 0.1$ &      $4.5 \pm 0.1$ &        $4.5 \pm 0.1$ \\
   \isotope[214]{Bi}, TPC &  $ \lesssim 800 $ &        $132 \pm 9$ &           $96 \pm 7$ \\
 \isotope[214]{Bi}, LXe shell &         $9 \pm 4$ &          $9 \pm 4$ &            $9 \pm 4$ \\
\end{tabular}

   \end{ruledtabular}
   }
 \end{center}
\end{table*}
Fig.~\ref{fig:xenon1t:0vbb:unblinded} shows the fit of the combined signal and background model to the unblinded data after the addition of the new MS cut with $\chi^2_\lambda/\text{ndf}=392/318=1.23$. Residuals below 2800 keV are symmetric around zero, while the model is mostly above the data points at higher energies. The parameter pulls are indicated by the black bars in Fig.~\ref{fig:xenon1t:0vbb:pulls}, and the best-fit numbers of background events around the Q-value are given in Tab.~\ref{tab:0vbb:backgrounds}.
For the \isotope[238]{U} and \isotope[232]{Th} chains, neither the trend nor the pulls are significantly changed compared to the blinded fit.

Contrary to the expectation the rate of \isotope[60]{Co} events is pulled close to zero.
The individual \isotope[60]{Co} peaks are present in the data outside of the fitting range and the best-fit components of the other backgrounds do not point to an overestimation of the acceptance for SS events from the \isotope[238]{U} and \isotope[232]{Th} decay chains.
This suggests that the strong pull is a feature of the stricter multi-scatter rejection.
With its double MeV-$\upgamma$ signature, \isotope[60]{Co} is different from the other background peaks, which do not feature secondary $\upgamma$-rays of equally high energies.
In order to detect the 2505.7~keV peak as an SS event, the $\upgamma$-rays need to be emitted in the same direction and fully absorbed within a few millimeters in $x$-$y$-$z$.
This makes an SS reconstruction of these events less likely than for the other background sources.
In the SS vs.~MS classification of MC events, only the $z$-separation of consecutive energy depositions was considered.
However, the MS selection on data also uses the S2 hit pattern which is sensitive to the $x$-$y$ separation of multiple scatters.
With the stricter post-unblinding cut, most \isotope[60]{Co} double-$\upgamma$ events were identified as MS and removed from the energy spectrum.
Since this was not modeled in the MC, the background model fit results in zero \isotope[60]{Co} events.

The tendency of the energy shift parameters is changed in the unblinded fit since the previously remaining MS events in the low-energy flanks of $\upgamma$-peaks biased the energy reconstruction. In the original fit, a stronger shift of the spectrum towards lower energies was observed which is not present after the removal of the events in the flanks. The best-fit cut acceptance at the Q-value is $85\,\%$. This is consistent with the acceptance-loss attributed to the new MS cut.

The standard deviation and mean position of the signal peak exhibit pulls close to zero.
Here, the fit is more sensitive to the constraint than to the data in absence of a signal.
The best-fit $\doublebeta$ decay rate is $(65\pm87)\;\text{t}^{-1}\text{yr}^{-1}$ which translates to $<210\;\text{t}^{-1}\text{yr}^{-1}$ at $90\,\%\;\text{CL}$.
From Eq.~(\ref{eq:0vbb:halflife:calc}), the lower limit on the half-life is 
\begin{equation}
 T_{1/2}^{\doublebeta} > 1.2 \times 10^{24}\;\text{yr}\; \text{at}\; 90\,\%\;\text{CL}.
\end{equation}
The resulting effective neutrino mass range, using NMEs from~\cite{Mustonen:2013zu, LopezVaquero:2013yji}, is $\langle m_{\upbeta\upbeta}\rangle =(0.8\text{–}2.5)\;\text{eV}/\text{c}^2$. 
Dedicated xenon-based \doublebeta{} experiments such as EXO-200 and KamLAND-Zen~\cite{exo200:PhysRevLett.123.161802, KamLAND-Zen:2022tow} have reported results that supersede our result by up to two orders of magnitude. In contrast to XENON1T, which is optimized for low background in the keV region, the dedicated detector designs are optimized to have low backgrounds at $Q_{\upbeta\upbeta}$. Together with \isotope[136]{Xe} enrichment, this leads to a more favorable signal-to-background ratio. The previously most stringent limit from DM direct detection experiments was set by PandaX-II with $T_{1/2}^{0\upnu\upbeta\upbeta}>2.3\times 10^{23}\;\text{yr}$ at $90\,\%\;\text{CL}$~\cite{PandaX-II:2019euf}. XENON1T improves on this result by an order of magnitude with lower background, larger exposure and four times better energy resolution. It illustrates the potential of current and future DM experiments such as LZ~\cite{Akerib:2019dgs}, XENONnT~\cite{xenonnt_mc:Aprile_2020}, DARWIN~\cite{DARWIN:2020jme} and beyond~\cite{Avasthi:2021lgy} for $\doublebeta$-decay searches. 

\section{Sensitivity of XENON\lowercase{n}T to $^{136}$Xe neutrinoless double-$\mathbf{\upbeta}$ decay}
\label{sec:nt_0vbb}
With its larger projected exposure, XENONnT will improve upon the XENON1T sensitivity to the \doublebeta{} process. This section discusses the sensitivity projections based on simulated XENONnT background data.

\subsection{Background model}
\label{subsec:nt_background}
The material backgrounds considered here originate from \isotope[60]{Co} as well as the \isotope[238]{U} and \isotope[232]{Th} decay chains.
The strategy to simulate the background events followed Sec.~\ref{mc:common}, but employed the XENONnT Geant4 geometry~\cite{xenonnt_mc:Aprile_2020}.
We also took the background contribution from the NV PMTs into account, which was at the same level as subdominant intrinsic backgrounds.

The intrinsic background sources considered for XENONnT include the daughter nuclei of \isotope[222]{Rn} isotopes emanated from the detector materials and the $\twonubeta$ decay of \isotope[136]{Xe} naturally present in the LXe, both discussed in Sec.~\ref{sec:bkg_1t}.
The expected \isotope[222]{Rn} contamination level of the LXe target is further reduced with respect to XENON1T with a novel online Rn removal system.
The \isotope[222]{Rn} activity concentration was assumed to be 1.0$\;\upmu\text{Bq}/\text{kg}$.

Due to the addition of 5.7~t of xenon to the total inventory, we conservatively assumed a natural abundance of \isotope[136]{Xe} in xenon, $\eta_{^{136}\text{Xe}}=(8.9\pm0.5)\times10^{-2}\;\text{mol}/\text{mol}$, using the difference with respect to the XENON1T measured abundance as a systematic uncertainty. The actual isotopic composition will be measured in the future. 

With a half-life of 3.82\;min and a Q-value at 4.17\;MeV, far beyond $Q_{\upbeta\upbeta}$, the $\upbeta$-decay of \isotope[137]{Xe} is a relevant background source in XENONnT.
It is produced through neutron capture on \isotope[136]{Xe} occurring either within the TPC itself or in the non-shielded regions -- outside the water tank -- of the experiment, especially in the LXe purification systems.
The mean travel time of xenon through the LXe purification system is approximately 7\;min, which is short enough for \isotope[137]{Xe} to be injected back into the detector before decaying.
We estimated a total \isotope[137]{Xe} production rate of $(6\pm5)\;\text{t}^{-1}\text{yr}^{-1}$ through a MC simulation.
Muon-induced neutrons produced in the LXe are primarily responsible for the production of \isotope[137]{Xe} in the TPC, contributing 10\% of the total rate.
The thermal neutron flux induced by radiogenic decays in rock, concrete, and materials is the dominant contribution to \isotope[137]{Xe} creation in the purification system, contributing 90\% of the overall rate.
The main source of uncertainty stems from different measurements of the thermal neutron flux at LNGS that are in tension with each other~\cite{neutron:best2016,neutron:belli1989,neutron:Rindi1988,neutron:Wulandari2004,Neutrons_fluxes:HAFFKE201136}. 

Neutrino-electron scattering is a potentially irreducible background source for the $\doublebeta{}$ decay search if the incident neutrino flux and energy are sufficiently high.
While the contribution from atmospheric neutrinos, diffuse supernova neutrinos, or geoneutrinos can be excluded, since either their flux or their energy is too small, the contribution from $^{8}$B solar neutrinos is relevant.
We used the neutral current \isotope[8]{B} neutrino flux measurement from the Sudbury Neutrino Observatory with $\Phi = (5.25 \pm 0.16_\text{stat} \pm 0.12_\text{sys})\times 10^{6}$ $\text{cm}^{-2}\;\text{s}^{-1}$~\cite{neutrino:Wulandari2013} and the $^{8}$B neutrino spectral shape from~\cite{neutrino:Winter2006} to derive the expected rate of neutrino-electron scatters in the detector, following the neutrino-electron elastic scattering cross-section calculation from~\cite{doi:10.1146/annurev-nucl-101918-023450}.
The electron neutrino survival probability follows the large mixing angle solution of the Mikheyev-Smirnov-Wolfenstein effect~\cite{Mikheyev:1985zog,PhysRevD.17.2369,Holanda_2003} which takes into account matter effects in the Sun.
\begin{figure}[t]
  \centering
  \includegraphics[width=\linewidth]{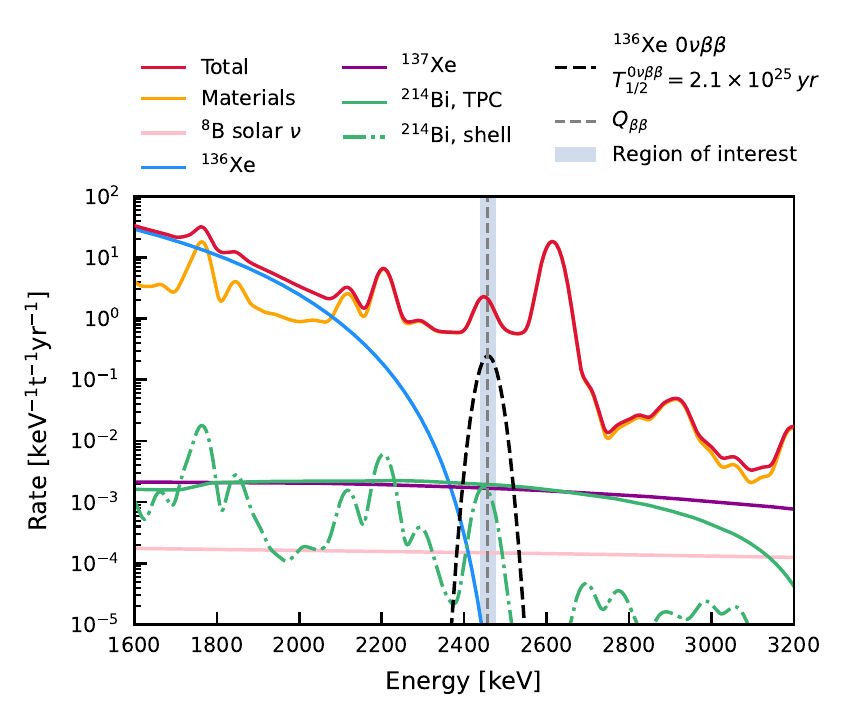}
  \caption{Energy spectrum of all backgrounds relevant for the \doublebeta-search in \nt.
  Dominant contributions around $Q_{\upbeta\upbeta}$ arise from material backgrounds (solid orange) and, in particular, from \isotope[214]{Bi} in the TPC (solid green), \isotope[137]{Xe} (solid purple) and the LXe shell (dash-dotted green). Backgrounds from $2\upnu\upbeta\upbeta$ of \isotope[136]{Xe} (solid blue) and \isotope[8]{B} solar neutrinos (solid pink) are subdominant.
  The shaded light blue area denotes the $2\,\sigma_E$ ROI.
  Event yields for the $2\,\sigma_E$ range around the Q-value are reported in Tab.~\ref{tab:constraints_nt}.}
  \label{fig:nT:er_bkg_summary}
\end{figure}

\subsection{Analysis}
\label{subsec:nt_analysis}
To maximize the sensitivity to the \doublebeta{} process, we employed a FV optimization analogous to Sec.~\ref{1t:ev_selection} in the $2\,\sigma_E$ region around $Q_{\upbeta\upbeta}$. We found an optimal FV with a mass of 1088~kg. The simulated energy spectra for all backgrounds within the FV are shown in Fig.~\ref{fig:nT:er_bkg_summary}.
The dominant background contribution around $Q_{\upbeta\upbeta}$ arises from the detector materials.
Each isotope considered in the MC simulation has a constraint term arising from the radioassay measurements, as indicated in Sec.~\ref{subsec:nt_background}, or from dedicated MC studies of the background. 
The fitting model was developed as described in Sec.~\ref{subsubsec:fit_blind}; the $\doublebeta$-signal was modeled as a Gaussian peak with an area proportional to the decay rate, $A_{\doublebeta}$.
\begin{table}[t]
\centering
  \begin{center}
  \caption{Expected background events in the 2$\sigma_E$ range around $Q_{\upbeta\upbeta}$ for 1000 days of live time and an FV mass of 1088~kg. There are no energy depositions from the early \isotope[238]{U} and \isotope[232]{Th} chains in the ROI. 
\label{tab:constraints_nt}}
  \begin{ruledtabular}
  \begin{tabular}{l c}
            Background source &   Expected Events \\
\colrule
            \isotope[238]{U} &             $ - $ \\
            \isotope[60]{Co} &   $ 0.3 \pm 0.2 $ \\
            \isotope[228]{Th} &     $ 70 \pm 40 $ \\
            \isotope[232]{Th} &             $ - $ \\
            \isotope[226]{Ra} &    $ 150 \pm 90 $ \\
            \isotope[214]{Bi}, LXe Shell & $ 0.16 \pm 0.08 $ \\
            \isotope[137]{Xe} &   $ 0.2 \pm 0.1 $ \\
            \isotope[214]{Bi}, TPC & $ 0.23 \pm 0.02 $ \\
            \isotope[8]{B} neutrinos &       $\leq 0.02$ \\
            \isotope[136]{Xe} \twonubeta &     $\leq 0.0005$ \\
\end{tabular}
  \end{ruledtabular}
  \end{center}
\end{table}

Differences to the XENON1T fitting model arise from additional backgrounds considered here and from omitting the nuisance parameters related to shifts of the peak positions.
Tab.~\ref{tab:constraints_nt} summarizes the expected event yields for an assumed live time of 1000 days in the $2\,\sigma_E$ region around $Q_{\upbeta\upbeta}$.
The dominant background contribution arises from the energy deposition of \isotope[214]{Bi} $\upbeta$- and $\upgamma$-emission where the $\upbeta$ was absorbed in the passive detector materials. The contribution of this decay product of the \isotope[226]{Ra} decay chain, is dominated by radioactive decays in the cryostat.

\begin{figure}[t]
  \centering
  \includegraphics[width=\linewidth]{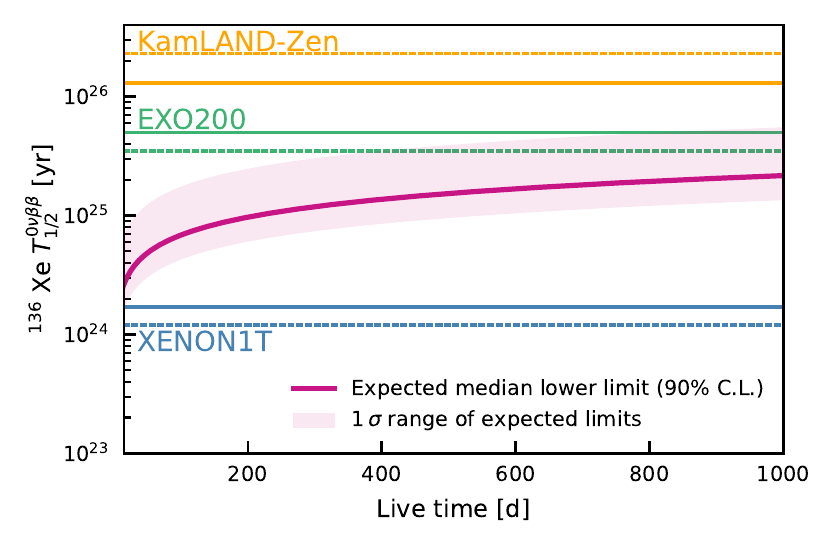}
  \caption{Expected median sensitivity for the lower limit on the half-life of \isotope[136]{Xe} $\doublebeta{}$ decay for XENONnT derived from Asimov data \cite{Cowan:2010js} with its $1\,\sigma$ statistical uncertainty. The projected sensitivity and the observed results from XENON1T, KamLAND-Zen~\cite{KamLAND-Zen:2022tow} and EXO-200~\cite{exo200:PhysRevLett.123.161802} are shown as solid and dashed lines, respectively.
  } \label{fig:nT:halflife_vs_lifetime}
\end{figure}

The physics reach of XENONnT was estimated with a profiled likelihood approach~\cite{Cowan:2010js} similar to the XENON1T analysis with the difference that we used the asymptotic assumption of test statistics whose validity was verified, see Sec.~\ref{fits:common}. We performed a likelihood scan for a set of assumed live times of the experiment (between 10 days and 1000 days) and determined the intersection with the $90\,\%$ quantile of a $\chi^2$-distribution with one degree of freedom.
This is the expected median lower limit that we report in Fig.~\ref{fig:nT:halflife_vs_lifetime}. The median lower limits for the considered live times were interpolated with a square-root function.

\subsection{Results}
\label{subsec:nt_results}
For a live time of 1000 days, we obtain a median lower limit of
\begin{equation}
  T_{1/2}^{\doublebeta} > 2.1 \times 10^{25}\;\text{yr}\; \text{at}\; 90\,\%\;\text{CL}.
\end{equation}
This is below the expected sensitivities and the observed lower limits from KamLAND-Zen and EXO-200~\cite{exo200:PhysRevLett.123.161802, KamLAND-Zen:2022tow} due to the large background contribution from detector materials and the low \isotope[136]{Xe} abundance.
Other LXe TPCs such as LZ~\cite{Akerib:2019dgs} or DARWIN~\cite{DARWIN:2020jme} are expected to be more sensitive due to a lower background level and larger FV mass, respectively. 

\begin{figure}[tpb!]
\centering
  \includegraphics[width=\linewidth]{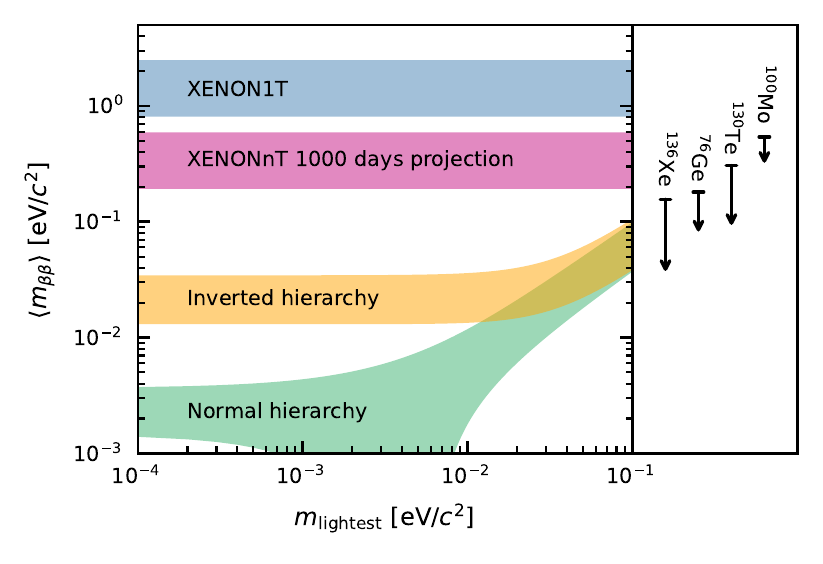}
  \caption{Effective Majorana neutrino  mass $\langle m_{\upbeta\upbeta} \rangle$ for XENONnT projection after 1000 days (violet), XENON1T (blue), and neutrino mass ordering depending on the mass of the lightest neutrino $m_{\textrm{lightest}}$.  The current best experimental limits for different double-$\upbeta$ candidate isotopes are shown in the right panel. 
  The values for \isotope[136]{Xe}, \isotope[76]{Ge}, \isotope[130]{Te}, and \isotope[100]{Mo} are taken from \cite{KamLAND-Zen:2022tow,PhysRevLett.125.252502,CUORE:2021mvw,CUPID:2020aow}, respectively.
  }
  \label{fig:nt_mbb_plot}
\end{figure}

From the derived lower limit of the half-life, we computed the effective Majorana neutrino mass $\mbb{}$ with the relation reported in Eq.~(\ref{eq:0vbb:effective_mass}).
We used the same assumptions regarding the NME and the phase-space factor as laid out in Sec.~\ref{subsec:postunblinding}.
We summarize our findings in Fig.~\ref{fig:nt_mbb_plot}, where the green band indicates the range of the effective Majorana neutrino masses for our XENONnT half-life sensitivity.
The masses range from $0.19\;\text{eV}/\text{c}^2$ to $0.59\;\text{eV}/\text{c}^2$ depending on the NME. While XENONnT is not yet competitive with dedicated experiments, this study shows that future xenon DM detectors can be competitive with optimized high-energy backgrounds and larger exposures.

\section{Conclusion and outlook}
\label{sec:conclusion}
In this paper, we reported on searches for double weak decays of $^{124}$Xe and $^{136}$Xe with XENON1T.
The search for 2$ \upnu\mathrm{ECEC}$ decay included a larger sample of data and an improved signal model compared to our previous result~\cite{XENON:2019dti}.
We detect $2\upnu\mathrm{ECEC}$ in \isotope[124]{Xe} with ${T_{1/2}^{\dec}=(1.1\pm0.2_\text{stat}\pm0.1_\text{sys})\times 10^{22}\;\text{yr}}$ at a significance of $7.0\,\sigma$. The half-life of this decay is the longest measured directly to date.

The search for $\doublebeta{}$ of \isotope[136]{Xe} is compatible with the background-only hypothesis with an exclusion limit of $T_{1/2}^{\doublebeta} > 1.2 \times 10^{24}\;\text{yr}\; \text{at}\; 90\,\%\;\text{CL}$.
Due to a larger active mass and expected lower background rate, the XENONnT experiment will improve this result.
With a live time of 1000 days, we expect a median lower limit of $T_{1/2}^{\doublebeta} > 2.1 \times 10^{25}\,\text{yr}\; \text{at}\; 90\,\%\;\text{CL}$.
While this is not competitive to dedicated searches, it demonstrates the feasibility of more sensitive searches in future xenon DM detectors.

The study of double-weak processes in LXe TPCs is not restricted to the two analyses discussed in this work and can be extended to a plethora of rare decays, such as the search for the $\twonubeta{}$ decay of $^{136}$Xe to the 0$^+_1$ excited state of $^{136}$Ba~\cite{Xe136:PhysRevC.93.035501}, the $\twonubeta{}$ and $\doublebeta{}$ decay of $^{134}$Xe~\cite{Xe134:PhysRevD.96.092001} or the neutrinoless second-order weak decays of $^{124}$Xe~\cite{Xe124:WITTWEG2020}.
Furthermore, a precise measurement of the $\twonubeta{}$ energy spectrum offers the possibility of experimentally testing the underlying nuclear models~\cite{SAAKYAN_2vbb_review,KamLandZen:PhysRevLett.122.192501}, but also to probe new physics beyond the SM~\cite{2vbbBSM:PhysRevLett.125.171801,2vbbBSM:PhysRevD.103.055019,2vbbBSM:AGOSTINI2021136127}.
The XENON project provides a broad science program ranging from DM searches to neutrino physics and properties of xenon, covering several orders of magnitude in energy.

\begin{acknowledgments}
We thank Dr.~J.~Kotila of Jyväskylä University for providing us with the $2\upnu\upbeta\upbeta$ SSD and HSD spectra. We gratefully acknowledge support from the National Science Foundation, Swiss National Science Foundation, German Ministry for Education and Research, Max Planck Gesellschaft, Deutsche Forschungsgemeinschaft, Helmholtz Association, Dutch Research Council (NWO), Weizmann Institute of Science, Israeli Science Foundation, Fundacao para a Ciencia e a Tecnologia, R\'egion des Pays de la Loire, Knut and Alice Wallenberg Foundation, Kavli Foundation, JSPS Kakenhi in Japan, Tsinghua University Initiative Scientific Research Program and Istituto Nazionale di Fisica Nucleare. This project has received funding/support from the European Union’s Horizon 2020 research and innovation programme under the Marie Sk\l{}odowska-Curie grant agreement No 860881-HIDDeN. Data processing is performed using infrastructures from the Open Science Grid, the European Grid Initiative and the Dutch national e-infrastructure with the support of SURF Cooperative. We are grateful to Laboratori Nazionali del Gran Sasso for hosting and supporting the XENON project.
\end{acknowledgments}

\bibliographystyle{spphys}
\bibliography{biblio}

\newpage
\onecolumngrid
\appendix*

\section{Tables with best-fit parameters and constraints}

\begin{table*}[h]
\centering
{
  \tabcolsep=1.0mm
  \caption{Best-fit signal and background model parameters for the XENON1T $2\upnu\text{ECEC}$ search that were shared among all datasets. Unitless parameters state the relative change to the expected value. The meaning of the parameters is given in Sec.~\ref{sec:edec:fit:method:pars}.}
  \label{tab:edec:fitpar_sr1_sr2_shared}
\begin{ruledtabular}
\begin{tabular}{lccc}
                              Parameter &       Fit value &      Constraint &                          Unit \\
\colrule
                $A_{2\upnu\text{ECEC}}$ &    $300 \pm 50$ &           $ - $ & $\text{t}^{-1}\text{yr}^{-1}$ \\
                          Solar $\upnu$ & $1.00 \pm 0.02$ & $1.00 \pm 0.02$ &                         $ - $ \\
                      \isotope[136]{Xe} & $0.99 \pm 0.03$ & $1.00 \pm 0.03$ &                         $ - $ \\
                       \isotope[238]{U} &   $1.0 \pm 0.5$ &   $1.0 \pm 0.6$ &                         $ - $ \\
                      \isotope[226]{Ra} &   $0.5 \pm 0.3$ &   $1.0 \pm 0.5$ &                         $ - $ \\
                      \isotope[232]{Th} &   $0.9 \pm 0.6$ &   $1.0 \pm 0.6$ &                         $ - $ \\
                      \isotope[228]{Th} &   $0.9 \pm 0.6$ &   $1.0 \pm 0.6$ &                         $ - $ \\
                       \isotope[60]{Co} &   $0.6 \pm 0.3$ &   $1.0 \pm 0.4$ &                         $ - $ \\
                        \isotope[40]{K} &   $1.0 \pm 0.3$ &   $1.0 \pm 0.3$ &                         $ - $ \\
   $\mu_{^{83\text{m}}\text{Kr,misID}}$ &  $32.1 \pm 0.6$ &  $32.1 \pm 0.6$ &                           keV \\
$\sigma_{^{83\text{m}}\text{Kr,misID}}$ &   $1.3 \pm 0.2$ &   $1.3 \pm 0.2$ &                           keV \\
     $f_{^{83\text{m}}\text{Kr,misID}}$ &   $2.6 \pm 0.4$ &   $2.6 \pm 0.4$ &                  $10 ^{ - 4}$ \\
\end{tabular}

\end{ruledtabular}}
\end{table*}

\begin{table*}[h]
\centering
{
 \tabcolsep=1.0mm
  \caption{Parameters of the combined signal and background model for the XENON1T $2\upnu\text{ECEC}$ search that were shared among science runs and fiducial volumes, fiducial volumes only, science runs only or that are exclusive to a single dataset (see column dataset). The meaning of the parameters is given in Sec.~\ref{sec:edec:fit:method:pars}.}
  \label{tab:edec:fitpar_sr1_sr2_exclusive}
\begin{ruledtabular}
\begin{tabular}{llccc}
                    Parameter &                    Dataset &                        Fit value &        Constraint &                          Unit \\
            $^{214}\text{Pb}$ &                        SR1 &                    $9.3 \pm 0.4$ &             $ - $ &    $\upmu\text{Bq}/\text{kg}$ \\
                              &                        SR2 &                    $5.3 \pm 0.8$ &             $ - $ &    $\upmu\text{Bq}/\text{kg}$ \\
             $^{85}\text{Kr}$ &                        SR1 &                    $0.7 \pm 0.1$ &     $0.7 \pm 0.1$ &                  $\text{ppt}$ \\
                              &                        SR2 &                    $1.2 \pm 0.2$ &     $1.2 \pm 0.2$ &                  $\text{ppt}$ \\
\colrule
            $^{125}\text{Xe}$ &             SR1$_\text{b}$ &                    $1.1 \pm 0.4$ &             $ - $ &    $\upmu\text{Bq}/\text{kg}$ \\
            $^{133}\text{Xe}$ &             SR1$_\text{a}$ &                  $0.00 \pm 0.01$ &             $ - $ &    $\upmu\text{Bq}/\text{kg}$ \\
                              &             SR1$_\text{b}$ &                  $1.24 \pm 0.03$ &             $ - $ &    $\upmu\text{Bq}/\text{kg}$ \\
                              &                        SR2 &                  $0.05 \pm 0.03$ &             $ - $ &    $\upmu\text{Bq}/\text{kg}$ \\
\colrule
         $A_{^{125}\text{I}}$ &             SR1$_\text{a}$ &                      $20 \pm 10$ &       $20 \pm 10$ & $\text{t}^{-1}\text{yr}^{-1}$ \\
                              &             SR1$_\text{b}$ &                     $570 \pm 90$ &     $700 \pm 100$ & $\text{t}^{-1}\text{yr}^{-1}$ \\
                              &                        SR2 &                      $50 \pm 30$ &       $40 \pm 30$ & $\text{t}^{-1}\text{yr}^{-1}$ \\
$A_{^{131\text{m}}\text{Xe}}$ &             SR1$_\text{a}$ &  $(2.4 \pm 0.1) \times 10^{ 3 }$ &             $ - $ & $\text{t}^{-1}\text{yr}^{-1}$ \\
                              &             SR1$_\text{b}$ &    $(156 \pm 1) \times 10^{ 3 }$ &             $ - $ & $\text{t}^{-1}\text{yr}^{-1}$ \\
                              &                        SR2 & $(21.4 \pm 0.6) \times 10^{ 3 }$ &             $ - $ & $\text{t}^{-1}\text{yr}^{-1}$ \\
\colrule
        $\kappa_\text{slope}$ & SR1$_\text{a,b}^\text{in}$ &                       $-6 \pm 2$ &        $-6 \pm 2$ &    $10^{-5}\,\text{keV}^{-1}$ \\
                              &  SR1$_\text{a}^\text{out}$ &                   $-1.8 \pm 0.4$ &    $-1.8 \pm 0.4$ &    $10^{-4}\,\text{keV}^{-1}$ \\
        $\kappa_\text{const}$ & SR1$_\text{a,b}^\text{in}$ &                $0.922 \pm 0.002$ & $0.922 \pm 0.002$ &                           $-$ \\
                              &  SR1$_\text{a}^\text{out}$ &                $0.914 \pm 0.003$ & $0.914 \pm 0.003$ &                           $-$ \\
                              &                        SR2 &                $0.931 \pm 0.001$ & $0.931 \pm 0.001$ &                           $-$ \\
 $A_{^{83\text{m}}\text{Kr}}$ &   SR1$_\text{a}^\text{in}$ &  $(3.5 \pm 0.1) \times 10^{ 3 }$ &             $ - $ & $\text{t}^{-1}\text{yr}^{-1}$ \\
                              &  SR1$_\text{a}^\text{out}$ &  $(3.4 \pm 0.1) \times 10^{ 3 }$ &             $ - $ & $\text{t}^{-1}\text{yr}^{-1}$ \\
                              &   SR1$_\text{b}^\text{in}$ &  $(6.9 \pm 0.2) \times 10^{ 3 }$ &             $ - $ & $\text{t}^{-1}\text{yr}^{-1}$ \\
                              &                        SR2 &  $(0.7 \pm 0.1) \times 10^{ 3 }$ &             $ - $ & $\text{t}^{-1}\text{yr}^{-1}$ \\
               $a_\text{res}$ &   SR1$_\text{a}^\text{in}$ &                $0.317 \pm 0.005$ & $0.313 \pm 0.007$ &    $\text{keV}^{\frac{1}{2}}$ \\
                              &  SR1$_\text{a}^\text{out}$ &                $0.315 \pm 0.006$ & $0.313 \pm 0.007$ &    $\text{keV}^{\frac{1}{2}}$ \\
                              &   SR1$_\text{b}^\text{in}$ &                $0.317 \pm 0.003$ & $0.313 \pm 0.007$ &    $\text{keV}^{\frac{1}{2}}$ \\
                              &                        SR2 &                $0.330 \pm 0.009$ &   $0.34 \pm 0.01$ &    $\text{keV}^{\frac{1}{2}}$ \\
               $b_\text{res}$ &   SR1$_\text{a}^\text{in}$ &                    $1.7 \pm 0.2$ &     $1.7 \pm 0.2$ &                     $10^{-3}$ \\
                              &  SR1$_\text{a}^\text{out}$ &                    $1.7 \pm 0.2$ &     $1.7 \pm 0.2$ &                     $10^{-3}$ \\
                              &   SR1$_\text{b}^\text{in}$ &                    $1.6 \pm 0.2$ &     $1.7 \pm 0.2$ &                     $10^{-3}$ \\
                              &                        SR2 &                   $-0.3 \pm 0.8$ &         $2 \pm 2$ &                     $10^{-3}$ \\
             $a_\text{shift}$ &   SR1$_\text{a}^\text{in}$ &                        $5 \pm 2$ &             $ - $ &                     $10^{-3}$ \\
                              &  SR1$_\text{a}^\text{out}$ &                       $-2 \pm 4$ &             $ - $ &                     $10^{-3}$ \\
                              &   SR1$_\text{b}^\text{in}$ &                    $1.8 \pm 0.6$ &             $ - $ &                     $10^{-3}$ \\
                              &                        SR2 &                    $2.5 \pm 0.4$ &             $ - $ &                     $10^{-2}$ \\
             $b_\text{shift}$ &   SR1$_\text{a}^\text{in}$ &                      $60 \pm 10$ &             $ - $ &                           keV \\
                              &  SR1$_\text{a}^\text{out}$ &                      $60 \pm 70$ &             $ - $ &                           keV \\
                              &   SR1$_\text{b}^\text{in}$ &                     $140 \pm 20$ &             $ - $ &                           keV \\
                              &                        SR2 &                      $90 \pm 10$ &             $ - $ &                           keV \\
\end{tabular}

  \end{ruledtabular}}
\end{table*}

\begin{table*}[h]
\centering
 \begin{center}
 \caption{Parameter constraints and best-fit parameters for the XENON1T $0\upnu\upbeta\upbeta$ search. Best-fit parameters are given for the blinded and unblinded data. Unitless parameters state the relative change to the expected value. For the acceptance, the constrained and best-fit acceptances are given instead of the fit parameters. Due to the implementation of the acceptance, the Gaussian constraint on the acceptance scaling parameter yields an asymmetric acceptance range. Due to the internal processing of the thorium chain, the respective best-fit parameters cannot be translated directly to $\isotope[228]{Th}$ event counts in Tab.~\ref{tab:0vbb:backgrounds}.}
   \label{tab:0vbb:constraints}
   {
   \tabcolsep=1.00mm
   \begin{ruledtabular}
\begin{tabular}{l c c c c}
                       Parameter &                           Unit &                      Constraints &                Blinded fit value &              Unblinded fit value \\
\colrule
                \isotope[238]{U} &                          $ - $ &                    $1.0 \pm 0.6$ &                    $0.6 \pm 0.1$ &                    $0.5 \pm 0.1$ \\
               \isotope[226]{Ra} &                          $ - $ &                    $1.0 \pm 0.5$ &                $0.620 \pm 0.008$ &                $0.624 \pm 0.008$ \\
               \isotope[232]{Th} &                          $ - $ &                    $1.0 \pm 0.6$ &                    $1.6 \pm 0.1$ &                    $1.8 \pm 0.1$ \\
               \isotope[228]{Th} &                          $ - $ &                    $1.0 \pm 0.6$ &                    $0.2 \pm 0.1$ &                    $0.0 \pm 0.1$ \\
                \isotope[60]{Co} &                          $ - $ &                    $1.0 \pm 0.4$ &                    $1.0 \pm 0.3$ &                   $-0.1 \pm 0.1$ \\
               \isotope[136]{Xe} &                          $ - $ &                  $1.00 \pm 0.03$ &                  $1.01 \pm 0.03$ &                  $1.01 \pm 0.03$ \\
          \isotope[214]{Bi}, TPC &     $\upmu\text{Bq}/\text{kg}$ &                            $ - $ &                    $1.7 \pm 0.1$ &                  $1.25 \pm 0.09$ \\
        \isotope[214]{Bi}, LXe shell &     $\upmu\text{Bq}/\text{kg}$ &                       $10 \pm 5$ &                       $10 \pm 5$ &                       $10 \pm 5$ \\
    $\Delta E\, _{\text{slope}}$ &                          $ - $ &  $(1.5 \pm 0.2) \times 10^{ 3 }$ &  $(1.5 \pm 0.1) \times 10^{ 3 }$ &  $(1.6 \pm 0.1) \times 10^{ 3 }$ \\
   $\Delta E\, _{\text{offset}}$ &                            keV &                   $-4.4 \pm 0.3$ &                   $-4.6 \pm 0.3$ &                   $-4.5 \pm 0.2$ \\
                      $\epsilon$ &                             \% &         $88.6^{ +8.9 }_{ -0.3 }$ &         $88.3^{ +0.6 }_{ -0.1 }$ &         $85.2^{ +4.6 }_{ -0.3 }$ \\
\colrule
    $\mu_{0\upnu\upbeta\upbeta}$ &                            keV &                 $2457.8 \pm 0.4$ &                            $ - $ &                 $2457.8 \pm 0.4$ \\
 $\sigma_{0\upnu\upbeta\upbeta}$ &                            keV &                   $19.7 \pm 0.3$ &                            $ - $ &                   $19.7 \pm 0.3$ \\
      $A_{0\upnu\upbeta\upbeta}$ &  $\text{t}^{-1}\text{yr}^{-1}$ &                            $ - $ &                            $ - $ &                      $60 \pm 90$ \\
\end{tabular}

     \end{ruledtabular}

  }
 \end{center}
\end{table*}

\end{document}